\documentstyle[preprint,aps]{revtex}
\begin{document}
\tightenlines
\draft
\title{The Charge and Matter radial distributions of Heavy-Light
mesons calculated on a lattice  with dynamical fermions.}
 
\author{UKQCD Collaboration}
\author{A.M. Green,
J. Koponen,
P. Pennanen\thanks{e-mails:anthony.green@helsinki.fi,
jmkopone@rock.helsinki.fi, petrus@leiki.fi} }
\address{Department of Physical Sciences and Helsinki Institute of Physics\\
P.O. Box 64, FIN--00014 University of Helsinki,Finland}
\author{C. Michael\thanks{e-mail: cmi@liv.ac.uk}}
\address{Department of Mathematical Sciences, University of Liverpool,
 L69 3BX, UK}

\date{\today}
\maketitle
\begin{abstract} A knowledge of the radial distributions of quarks
inside hadrons could lead to a better understanding of the QCD
description of these hadrons and possibly suggest forms
for phenomenological models. As a step in this direction, 
in an earlier work, the charge (vector) and matter (scalar) radial
distributions of  heavy-light mesons were measured in the quenched
approximation on a
$16^3\times 24$ lattice with a lattice spacing of 
$a\approx 0.17$~fm, and a hopping parameter
corresponding to a light quark mass about that of the strange quark.

Here several improvements are now made: 
1) The configurations are generated using dynamical
fermions with $a\approx 0.14$~fm; 
2) Many more gauge configurations are
included; 3) The distributions at many off-axis, in addition to on-axis, 
points are measured;
4) The data analysis is much more complete. In particular, distributions
involving excited states are extracted.

The exponential decay of the charge and matter distributions can be
described by mesons of mass 0.9$\pm 0.1$ and 1.5$\pm 0.1$
GeV respectively --- values that are consistent with those
of vector and scalar $q\bar{q}$-states calculated {\em directly} with the same
lattice parameters.
 \end{abstract}
 
\newpage
\section{Introduction}
 \label{intro}
In few- and many-body systems radial and momentum distributions often play
an important role. For atomic and nuclear systems these distributions
are, in many cases, calculated from a differential equation using
an effective interparticle potential, where both the equation and potential
have some justification. However, for quark-gluon systems this approach  
is thought not to be applicable, even though the basic interaction ---
that of QCD --- is exact and well known. Therefore, if --- for example ---
transition rates between  states in a heavy-light meson ($Q\bar{q}$)
are calculated, then the necessary radial wavefunctions are simply taken to have
some convenient form, as in Ref.~\cite{Iach}, or are calculated with a
differential equation and interquark interaction that are not well
justified.
It is an unusual situation, where one of the most fundamental
systems,  the hydrogen atom of quark physics,  has interparticle
correlations that are little understood.

In an attempt to remedy partially 
this problem, the authors in Ref.~\cite{G+K+P+M} {\em measured} the charge
and matter distributions in a heavy-light meson. More explicitly, the 
heavy-light meson was simplified to being an infinitely heavy quark ($Q$)
 and an antiquark ($\bar{q}$) with a mass approximately equal to that of
the strange quark. The physical meson nearest to this idealised meson
is the $B_{{\rm s}}$(5.37 GeV). Ref.~\cite{G+K+P+M} was essentially a pilot
calculation to test the feasibility of such distribution measurements
and its successful outcome encouraged the authors to continue this line
of research. In this paper, the same correlations are measured but with
several major improvements compared with the earlier study:

  1) The gauge configurations are now calculated using  two flavors of
dynamical fermion
and not in the quenched approximation as before. The
actual parameters are those in Ref.~\cite{Allton}, namely, $\beta=5.2$,
$C_{{\rm SW}}=1.76$ for the clover action, $a\approx 0.14$~fm for the
lattice spacing, $\kappa=0.1395$ for the hopping parameter and $M_{{\rm
PS}}/M_V=0.72$ for the pseudoscalar meson/vector meson mass ratio, 
which corresponds  to a quark mass somewhat heavier than the strange quark
mass --- $M_{{\rm PS}}/M_V=0.682$ being the ratio corresponding to
exactly the strange quark mass\cite{qmass}.
In comparison, the parameters for the quenched work of
Ref.~\cite{G+K+P+M} were $\beta=5.7$, $C_{{\rm SW}}=1.57$, $a\approx
0.17$~fm, $\kappa=0.14077$ and $M_{{\rm PS}}/M_{{\rm V}}=0.65$,
corresponding to~\cite{MP98} a  value of $m_{\bar{q}}=0.91(2) m_{{\rm
s}}$. The spatial lattice size is
2.24~fm (compared to 2.72~fm in the
previous quenched calculation).

2) Many more gauge configurations are generated --- 78 with dynamical
fermions compared with the earlier 20 quenched configurations. 

3) Previously the densities were measured only at the 7 on-axis
points $r=0, 1,\ldots,6$. Now the densities at 20 points are measured ---
the same 7 on-axis with the addition of 13 off-axis points. This
permits a potentially much more detailed mapping of the density
profiles. Furthermore, it opens up the possibility of testing
the rotational invariance of these profiles by comparing them
at $r=5$ with that at $(x=3, \ y=4)$ and also $r=3$ with $(x=2, \ y=2, \ z=1)$.
This symbolic notation for an off-axis
point will be used throughout the article. In practice, it includes
the 8 possibilities $(x=\pm 3, y=\pm 4)$ and $(x=\pm 4, y=\pm 3)$, 
so that when combined with 
the  directions along the 3 axes, this makes 24 independent
measurements for each symbolic $(x, \ y)$, when $x\not= y$. In contrast,
the on-axis cases have only 6 independent measurements for
each $r$. This will be seen to improve significantly  the statistics for
the off-axis points compared with their neighboring on-axis points. This is
in spite of the fact that an off-axis point requires a longer string of
latticized links, so that ---  being guided by strong coupling arguments ---
it should be more difficult to measure.

4) Since the work in Ref.~\cite{G+K+P+M}, the methods and
our understanding of the data analysis have been developed. In
particular, the interesting off-diagonal density terms are always 
allowed to vary and are no longer fixed to zero as was sometimes 
the case earlier. This now gives a better estimate of excited state
effects. Also for the radial dependence of the density $x^{\alpha\beta}(r)$, 
the use of a separable form $y_{\alpha}(r)y_{\beta}(r)$ is also
employed. Here $\alpha, \  \beta$ are state indices with $\alpha =1$
being the ground state. This separable form
is found to have some interesting features not found in the
non-separable approach. Furthermore, {\em if }  the densities were to be
interpreted in terms of  underlying wavefunctions $\psi_{\alpha}(r)$, 
then the separable form
$x^{\alpha\beta}(r)=\psi_{\alpha}(r)\psi_{\beta}(r)$
would be perfectly natural. However, it should be emphasised that such
an interpretation can only have a phenomenological justification.

In Section~\ref{formalism}  the two- and three-point
correlation functions needed to extract the densities  are 
briefly discussed --- the reader being referred to Ref.~\cite{G+K+P+M}
for more details. In Section~\ref{analysis} the methods for analysing
the basic lattice data are described. This results in values for the
ground and some excited state energies and, in addition, radial
distributions of the charge (vector) and matter (scalar) densities for
these states are extracted.
 In Section~\ref{radialdep} these radial distributions    are
parametrized in terms of latticized forms of Yukawa, exponential and
gaussian functions.
In Section~\ref{con} a summary and some conclusions are made.

\section{The correlation functions $C_2$ and $C_3(r)$}   
\label{formalism}

\begin{figure}[ht]
\includegraphics{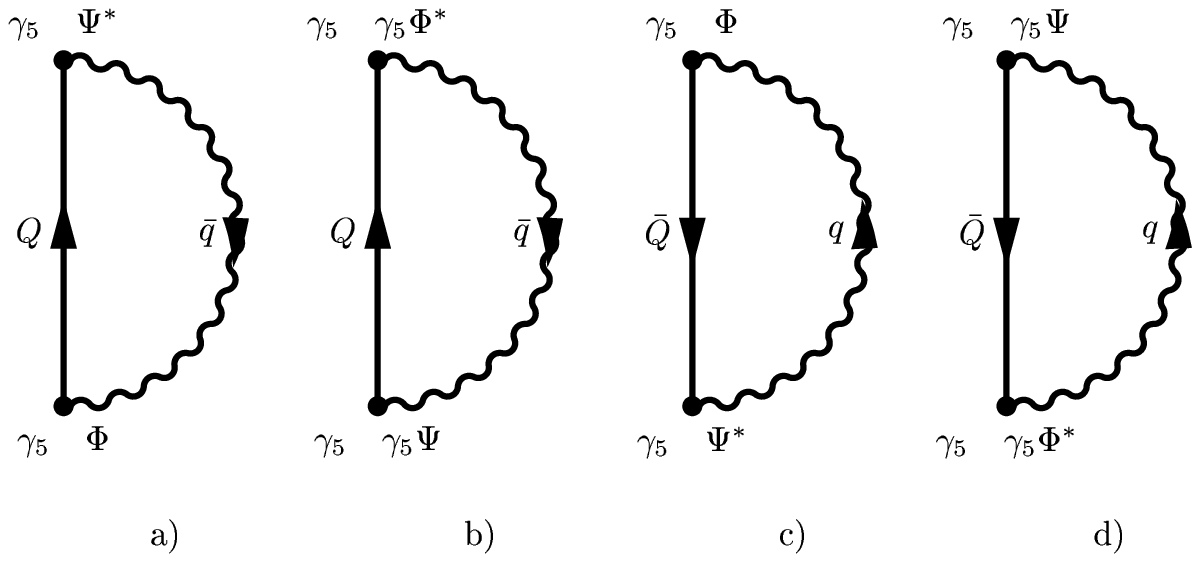}
\caption{The four contributions to the two-point correlation function
$C_2$} 
\label{c2diags}
\end{figure}

In this work the  basic entities are the two- and three-point
correlations [$C_2, \ C_3$],  both of which are needed for
measuring radial distributions. These are depicted 
in Figures \ref{c2diags} and \ref{c3diags} and are seen to be
constructed from essentially two  quantities --- the heavy (static)-quark
propagator $G_Q$ and the light-quark propagator $G_q$.

As discussed in detail in Ref.~\cite{G+K+P+M},
when the
heavy-quark propagates from site $({\bf x}, \ t)$ to site $({\bf
x'}, \ t+T)$,  $G_Q$ can be expressed as
 \begin{equation}
\label{GQ}
G_Q({\bf x}, \ t \ ; \ {\bf x'}, \ t+T)=
\frac{1}{2}(1+\gamma_4) U^Q({\bf x},t,T)\delta_{{\bf x},{\bf x'}} ,
 \end{equation}
 where $U^Q({\bf x},t,T)=\prod^{T-1}_{i=0} U_4({\bf x},t+i)$ is the
gauge link product in the time direction.
On the other hand, as  the light-quark
propagates   from  site $i$ to site $j$, it can be schematically
 expressed as~\cite{MP98}   
 \begin{equation}
\label{Gauss2}
G_q=G_{ji}= Q^{-1}_{ji}=\langle (Q_{ik}\phi_k)^*\phi_j\rangle=
\langle \psi^*_i\phi_j\rangle. 
\end{equation}
or as
 \begin{equation}
\label{Gauss3}
G_q'= G'_{ji}=\gamma_5\langle (Q_{jk}\phi_k)\phi_i^*\rangle\gamma_5=
\gamma_5\langle\psi_j\phi_i^*\rangle\gamma_5. 
\end{equation}
Here the $\phi_i$ are pseudo-fermions situated on the lattice sites $i$
and $\psi_i=Q_{ik}\phi_k$, where $Q$ is the Clover-Wilson-Dirac
matrix, which is specified by $C_{\rm {SW}}$ and  the hopping parameter $\kappa$.

Knowing $G_Q$ and $G_q$, 
the general form of a two-point correlation can be constructed from a 
heavy-quark propagating from site 
$({\bf x}, \ t)$ to site $({\bf x'}, \ t+T)$ and a light-quark 
propagating from site 
$({\bf x'}, \ t+T)$ to site $({\bf x}, \ t)$ as
\begin{eqnarray}
\label{C_2}
C_2(T)=&\hspace{-17mm}
{\rm Tr} \langle \Gamma^{\dagger}\, G_Q({\bf x},t;{\bf x'},t+T)
\, \Gamma \, G_q({\bf x'},t+T;{\bf x},t)\rangle\nonumber\\
=& 2 \langle
{\rm Re }\left[U^Q [\psi^*({\bf x},t+T)\phi({\bf x},t)+
\phi^*({\bf x},t+T)\psi({\bf x},t)]\right] \rangle.
\end{eqnarray}
Here $\Gamma$ is the spin structure of the heavy-quark light-quark
vertices at $t$ and $t+T$. In this case $\Gamma=\gamma_5$, since we are 
only interested in pseudoscalar mesons such as the $B$-meson.  For
clarity, the Dirac indices have been omitted.

Similarly, when the light-quark field is probed by an operator
$\Theta( {\bf r})$ at $t=0$ and the heavy-quark propagates from $t=-t_2$
to $t=t_1$
\begin{equation}
\label{AC_3Q}
 C_3(-t_2, \ t_1, \ {\bf r})={\rm Tr} \langle \Gamma^{\dagger} \ G_Q({\bf
x}, \ -t_2 \ ;{\bf x}, \ t_1) \Gamma \ G_q({\bf x}, \ t_1; {\bf x+r},\
0) \ \Theta ({\bf r}) \   G'_q({\bf x+r},\ 0; {\bf x}, \ -t_2) \rangle.
 \end{equation}
Here $\Theta=\gamma_4$ for the charge distribution and 1 for the 
matter (scalar) density.

The above  has been written down for a single type of gauge field.
However, the correlations can be greatly improved by fuzzing. 
 In this case the basic links containing the gauge field have two
fuzzings in addition to the original
 local field $(L)$.  In the standard notation of, for example
Ref.~\cite{GMP93},  Fuzz1 has 2 iterations and Fuzz2 a further 6
iterations i.e. 8 in all.
These will be referred to as $F_1$ and $F_2$. In both cases, the factor
multiplying the basic link is $f_p=2.5$ i.e.

 [Projected fuzzed link] = $f_p\cdot $[Straight link]+ [Sum of 4 spatial
U-bends]

\noindent with the quarks separated by a product of  fuzzed
links of length 1 lattice unit for Fuzz1 and 2 lattice units for Fuzz2, 
as discussed in Ref.~\cite{G+K+P+M}.
With fuzzing included, $C_2$ and $C_3$ are now $3\times 3$ matrices
composed of matrix elements with the indices  
LL, LF$_1$, LF$_2$, F$_1$F$_1$, F$_1$F$_2$ and F$_2$F$_2$.
This means that S-wave excited state energies and properties can now be studied
in addition to those of the ground state.

\section{ Analysis} 
\label{analysis} 
 
There are several ways of analysing the above correlation functions
$C_2$ and $C_3$ in order to extract the quantities of interest i.e.
energies and radial distributions. For a review of these methods see
Ref.~\cite{McN+M} --- with more details using the present notation  being
found  in Ref.~\cite{G+K+P+M}. We now draw upon 
experiences learnt in that reference.

Firstly, the two-point correlation data $C_2$  are analysed to give the
energies ($m_{\alpha})$ and eigenvectors $({\bf v})$ for the states of the
$Q\bar{q}$-system. These values of $m_{\alpha}$ and  ${\bf v}$ are 
then fixed when analysing the three-point
correlation  data $C_3$ to give the charge and matter densities
$x^{\alpha \beta}(r)$.
 
\subsection{Analysis of the two-point correlation functions $C_2$}

Consider the  correlation function $C_2(T)$  as an $n\times n$ matrix --- 
$3\times 3$ in this case with the elements LL, LF$_1$, $\ldots$, F$_2$F$_2$. Each
element $C_{2,ij}(T)$ is then fitted with the form
 \begin{equation}
\label{Cij}
C_{2,ij}(T)\approx \tilde{C}_{2,ij}(T)=\sum_{\alpha =1}^{M_2}v_i^{\alpha}
\exp(-m_{\alpha}T)v_j^{\alpha},
\end{equation}
where $M_2$ is the number of eigenvalues and   $m_1$ is the ground state
energy of the heavy-light meson. 
The values of $m_{\alpha}$ and $v_{i,j}^{\alpha}$ are then determined  
by  minimizing the
difference between the $C_2$ data from the lattice and  the form 
$\tilde{C}_2$. The function actually minimized is the usual
\begin{equation}
\label{chim}
\chi^2=\sum_{i,j} \sum_{T_{2,{\rm min}}}^{T_{2,{\rm max}}}
\left[\frac{C_{2,ij}(T)- \tilde{C}_{2,ij}(T)}{\Delta C_{2,ij}(T)}\right]^2,
\end{equation} 
where $\Delta C_{2,ij}(T)$ is the statistical error on $C_{2,ij}(T)$
and $T_{2,{\rm min}}, \ T_{2,{\rm max}}$ are the minimum and maximum values  of 
$T_2$ used in the fit. Here $T_{2,{\rm max}}$ is fixed at 11 --- the maximum
possible value when applying the Maximum Variance Reduction  method on a
$16^3\times 24$ lattice --- see Ref.~\cite{MP98}.
The minimization is carried out using the program package {\sf Minuit} --- the
{\sf Migrad} option being the most successful, since  this enables errors on the
varied parameters $m_{\alpha}$ and $v_{i,j}^{\alpha}$ to be determined.  
It was unnecessary  to use the {\sf Minos} option, since the errors were
found to be quite symmetrical.

Before the above minimization is started,  two parameters  need 
to be decided upon --- $M_2$ and  $T_{2,{\rm min}}$.
The criteria for this choice are that the $\chi^2/n_{{\rm dof}}$ must be
small --- to be satisfied that the fit to the data is good ---  and 
that $T_{2,{\rm min}}$ is appropriate for  future use of the
$m_{\alpha}$ and $v_{i,j}^{\alpha}$ in analysing the $C_3$ data. 
The latter data covers a range from $T_3=4$ to 10,
where each $C_3$ is constructed from essentially the product of 
two $C_2$'s i.e.  $C_3(T_3=10)$ requires a good fit to
$C_2(T_2=5)$, whereas  $C_3(T_3=4)$ would require a good fit to
$C_2(T_2=2)$. However, it must be kept in mind that ultimately,
when extracting densities from $C_3(T_3)$, it is the larger
values of $T_3$ --- where convergence hopefully occurs --- that 
are of more importance. In practice, this means $T_{3,{\rm min}}\approx 8$
is a reasonable compromise, so that the $C_2(T_2)$ fit should
concentrate on values of $T_2\ge 3$.

In Ref.~\cite{G+K+P+M}, $M_2= 2, 3$ and $T_{2,{\rm min}}= 3, 4$ were considered.
However, the $M_2=2$ and $T_{2,{\rm min}}= 3$ fits resulted in considerably 
larger   $\chi^2/n_{{\rm dof}}$ and were not used in the subsequent $C_3$ 
analysis. Here the situation is similar with now $M_2=3,4$ and
 $T_{2,{\rm min}}= 3, 4$ giving the best and most appropriate fits to $C_2$. 
The outcome is shown in Table~\ref{Tablec2fit}, where we present 4 cases.
In most of this paper
 we will concentrate on Case B, since this has both a good $\chi^2/n_{{\rm dof}}$
of 0.16 and sufficiently small  errors on the state energies $am_{\alpha}$. 
In contrast, Case A with $M_2=3$ has a large $\chi^2/n_{{\rm dof}}$ and
Case C large errors on the $am_{\alpha}$.  

The table also shows --- with
Case Q --- the earlier best fit in Ref.~\cite{G+K+P+M} to the 20 quenched
configurations with $\beta=5.7$. When comparing the $am_{\alpha}$ from
the four cases, two points must be kept in mind:
 
\noindent 1) Only differences of
the $am_{\alpha}$'s have a meaning, since the lattice simulation
generates different self-energies to the quarks in Case Q versus Cases
A, B, C. 

\noindent 2) The table shows $am_{\alpha}$, where the lattice spacing
$a$ is $\approx 0.17$~fm for Case Q and  $\approx 0.14$~fm for Cases A, B, C.

\noindent Removing these two effects results in the 
$\Delta m_{\beta \alpha}=am_{\beta}-am_{\alpha}$ at the bottom of 
Table~\ref{Tablec2fit}. There it is seen that, within the error bars,
both $\Delta m_{21}$ and $\Delta m_{31}$ are unchanged in going from the
20 quenched configurations --- after being scaled by the ratio of the
lattice spacings 0.14/0.17 ---
to the preferred unquenched Case B. The best
estimates are $\Delta m_{21}=0.33(1)$, $\Delta m_{31}=0.80(4)$ and 
$\Delta m_{41}=1.05(6)$. This value of  $\Delta m_{21}$ is also the same as
that obtained in Ref.~\cite{MP98}, when the latter is also scaled by the
lattice spacing ratio 0.14/0.17. However, it should be pointed out that
in Ref.~\cite{MP98} the quenched approximation was used with the same 
parameters as in Ref.~\cite{G+K+P+M} but on a $12^3$ spatial lattice.

In Table~\ref{Tablec2fit}, for Case B, there are 54 bits of data fitted
with 16 parameters. However, it should be added that the Parameter
Correlation Coefficient from the {\sf Migrad} algorithm of {\sf Minuit} indicates
that several of these
parameters are correlated. For example, there is a strong correlation
between $am_1$ and $v^1_{{\rm L}}$ and, if $v^1_{{\rm L}}$ is fixed
as $v^1_{{\rm L}}=0.45046\cdot am_1$,
then {\sf Migrad} finds the same solution. In this way 9 of the 16
parameters can be effectively removed to leave only 7 --- the 4
$m_i$ and $v^1_{{\rm F}_1,\ {\rm F}_2}$, $v^3_{{\rm F}_2}$ --- but
with the same final solution.
Therefore, the  $n_{2,{\rm dof}}$ increases from 38 to 47 and the 
$\chi^2/n_{2,{\rm dof}}$ decreases to 0.13. Such small values of 
$\chi^2/n_{2,{\rm dof}}$ are not surprising, since also correlations within
the data are expected. These are of three forms:\\
 1) For a given $C_{ij}(T)$, there may be correlations between the 
78 gauge configurations. However, this has been   greatly reduced 
selecting  the 78 configurations used to be separated by
40 trajectories. 
This feature can be checked 
by calculating the autocorrelation function ($f$) between the $N$=78 
configurations to give an effective number of configurations
$N_{{\rm effective}}=N/f$. Here it is found that $f\approx 1$ --- 
indicating a negligible correlation.\\
 2) For a given gauge configuration, there are correlations
between the different $T$-values for both $C_2(T)$ and $C_3(T)$. This
has been studied in, for example, Ref.~\cite{JL}, where the general
conclusion was that consideration of these types of correlation
resulted in essentially the same final eigenvalues but, in some cases,
 with somewhat increased error bars.\\
3) There could be correlations between the different values of $r$.
Hopefully, our employment of various analysis schemes --- such as the use of
several different values of $T_{{\rm min}}$ and also the non-separable versus
separable forms discussed below --- will minimize the uncertainty in this
particular correlation.

\subsection{Analysis of the three-point correlation function  
for radial distributions}
\label{A3pf}
The analysis of the three-point correlation functions $C_3(\Theta,T,r)$
is performed using a generalisation of the one for $C_2$ in 
Eq.~\ref{Cij}, namely,
 \begin{equation}
\label{C3fit}
C_{3,ij}(\Theta ,T,r)\approx \tilde{C}_{3,ij}(\Theta ,T,r)=
\sum_{\alpha =1}^{M_3}\sum_{\beta =1}^{M_3}v_i^{\alpha}
\exp[-m_{\alpha}t_1]x^{\alpha\beta}(r)\exp[-m_{\beta}(T-t_1)]v_j^{\beta}.
 \end{equation}  
The $m_{\alpha}$ and ${\bf v}$-vectors are those obtained by 
minimizing the $C_2$ in Eq.~\ref{Cij}
and, for each value of~$r$, the $x^{\alpha \beta}(r)$
are varied to ensure a good fit to $C_{3,ij}(\Theta,T,r)$ by the model
form $\tilde{C}_{3,ij}(\Theta ,T,r)$.
As for $C_2$ a decision must be made on the maximum and minimum
values  $T_{3,{\rm max}}, \ T_{3,{\rm min}}$ of $T_3$.
Now  $T_{3,{\rm max}}$ is fixed
at 10, since the signal to noise ratio was too large for larger values
of T. As discussed above, when deciding on
$T_{2,{\rm min}}$, we consider $T_{3,{\rm min}}\approx$ 6--9
to be a reasonable range.

Two forms of  $x^{\alpha \beta}(r)$ are used here:\\
\noindent 1) A non-separable (NS) form, where each $x^{\alpha \beta}(r)$ is treated as
a single entity. Here we take $M_3=M_2=4$. However, for {\sf Migrad} to converge
to a reasonable solution, of the 10 possible values of $x^{\alpha \beta}(r)$
for a given value of $r$, only
7 are varied --- the 4 $x^{\alpha \alpha}(r)$ and the 3 $x^{1 \alpha}(r)$
with $\alpha \not=1$. The other $x^{\alpha \beta}(r)$ are fixed to be zero.
As discussed above for $C_2$, {\sf Migrad} indicates correlations between
these 7 variables. However, taking into account
these correlations  results in the same solution.\\
\noindent 2) A separable (S) form 
$x^{\alpha \beta}(r)=y_{\alpha}(r) y_{\beta}(r)$.
Here, we take $M_3=3$ to give only three free parameters for each value
of $r$ --- $y_1(r)$, $y_2(r)$ and $y_3(r)$.

\noindent  It is seen from Section~\ref{formalism} that this second
form appears to be a more natural parametrization,
since schematically we can write $C_3$  as
$\langle \gamma_5 G_Q(-t_2\rightarrow t_1)\gamma_5 G_q(t_1\rightarrow 0)
\Theta G_q(0\rightarrow -t_2)\rangle $  in a form containing two light-quark 
propagators. 
Because these propagators act over different time intervals, $C_{3,ij}$ 
might  --- to some extent --- be separable in terms of the indices 
$\alpha$ and $\beta$.  
As shown in the Appendix, when the separable form $x^{\alpha
\beta}(r)=y_{\alpha}(r) y_{\beta}(r)$  is used in Eq.~\ref{C3fit}
the ratio
 \begin{equation}
\label{ratio}
R_{ij}(T)=\ln \left[ \frac{\tilde{C}_{3,ij}(T-1)}{\tilde{C}_{3,ij}(T)}\right]
 \end{equation}  
has, when $T$ is odd, the symmetry
 \begin{equation}
\label{sym}
R_{ij}(T)=R_{ij}(T+1).
 \end{equation}  
The surprising point is that this {\em same} symmetry is also found, to
a greater or lesser extent, in the basic data. This is shown in
Table~\ref{symt} for the dominant correlation $C_{3,{\rm F}_1{\rm F}_1}$
at the small interquark distances  of $r=1$ and the off-axis points 
$(x=1, \ y=1)$ and $(x=1, \ y=1, \ z=1)$ the symmetry is very clear.
However, as $r$ increases the symmetry essentially disappears --- as
seen in the last column for $(x=3, \ y=4)$.

In Table~\ref{Tablechden} results are given for both the NS and S forms
of $x^{\alpha \beta}(r$) and for different choices of $T_{2,{\rm min}}$,
$T_{3,{\rm min}}$
--- with the most representative  solution being $S(3,8)$, the
separable form with $T_{2,{\rm min}}=3$ and $T_{3,{\rm min}}=8$.
The other choices give  support to this solution and indicate the
possible systematic error.  In the penultimate column is given, in our
opinion, the best overall estimate of the ground state charge density
with error. The last column shows
estimates using Case A in Table~\ref{Tablec2fit}. As shown in 
Ref.\cite{G+K+P+M}, in this case the $v_i^{\alpha}$ matrix is square
and so can be inverted to give the matrix $u_i^{\alpha}$. Estimates
$\bar{C}_{3,\alpha \beta}(T)$ can then be written down directly as
\begin{equation}
\label{barCij}
\bar{C}_{3,\alpha \beta}(T)=u_i^{\alpha}C_{3,ij}(T)u_j^{\beta}.
\end{equation}
 In this case
\begin{equation}
\label{xij}
x^{11}(r)=\lim_{T\rightarrow \infty}
\frac{\langle \bar{C}_{3,11}(T,r)\rangle}{\langle \bar{C}_{2,11}(T)\rangle}.
\end{equation}
Unfortunately, the extraction of the asymptotic 
$T\rightarrow \infty$ limit is somewhat
subjective and gives the estimates in the last column in 
Table~\ref{Tablechden}. In all  the cases listed, within errors these agree
with the previous column. However, this approach did show that the data from
some of the larger values of $(x,y)$ were not good and so these  are dropped in
the subsequent discussion. 
Also, for reasons to be discussed in Section~\ref{RotationI}, neither the
(2,2,1) data nor the weighted average of (3,0,0)/(2,2,1) are included in
the following analysis.

 In Table~\ref{Tablechdenex} 
similar  estimates are given for charge densities involving excited
states. Also this table contains a summary of the $x^{11}(r)$ and 
$x^{12}(r)$ matter radial
distributions, which were extracted using both the NS and S forms for
different choices of $T_{2,{\rm min}}$, $T_{3,{\rm min}}$ 
--- just as in the charge case.
However, these signals are somewhat weaker than for the charge, so that
no meaningful matter distributions could be extracted for $r\ge 4a$.

In Figure~\ref{x11vs2LY} the best estimates of the charge and
matter distributions from 
Tables~\ref{Tablechden} and \ref{Tablechdenex} are compared.
To guide the eye we also show lines depicting 
the lattice exponential fits to be discussed later in Section~\ref{CSF}. 
Here it is clearly seen that the range of the charge distribution
is longer than that of the matter distribution.
Furthermore,  this figure also contains the
charge and matter densities  obtained with the quenched approximation in
Ref.~\cite{G+K+P+M}. For this  comparison, the results of 
Ref.~\cite{G+K+P+M} are scaled from lattice spacing $a_{0.14}$ to
$a_{0.17}$ by simply $r_{0.17}\rightarrow \rho r_{0.17}$ and
$x^{11}(r_{0.17})\rightarrow \rho^{-3}x^{11}(r_{0.17})$, where 
$\rho=0.17/0.14$.
Here it is seen that the present results using dynamical fermions are
indistinguishable from those using the quenched approximation. 
It will be seen later in the last two columns
of Table~\ref{Chsumrule}, that this  near equality is also 
reflected  in the charge sum rules with both giving
$x^{11}=1.4(1)$.
As discussed in Refs.~\cite{foster} and \cite{Fosterthesis}, this is of 
particular interest in the matter case, since there disconnected
contributions could arise that are dependent on the quenched versus
unquenched. Any difference would then be due to the effect of the quark 
condensate. Clearly, with the present data no such effect can be detected. 
However, it must be remembered that here the sea quarks have the same mass as
the valence quarks i.e. about that of the strange quark. It is possible
that using sea quarks with $u,d$ masses the above conclusion would be 
different. 
This observation that full QCD and the quenched approximation
do not differ significantly was has been seen many times before.

\begin{figure}[ht] 
\includegraphics{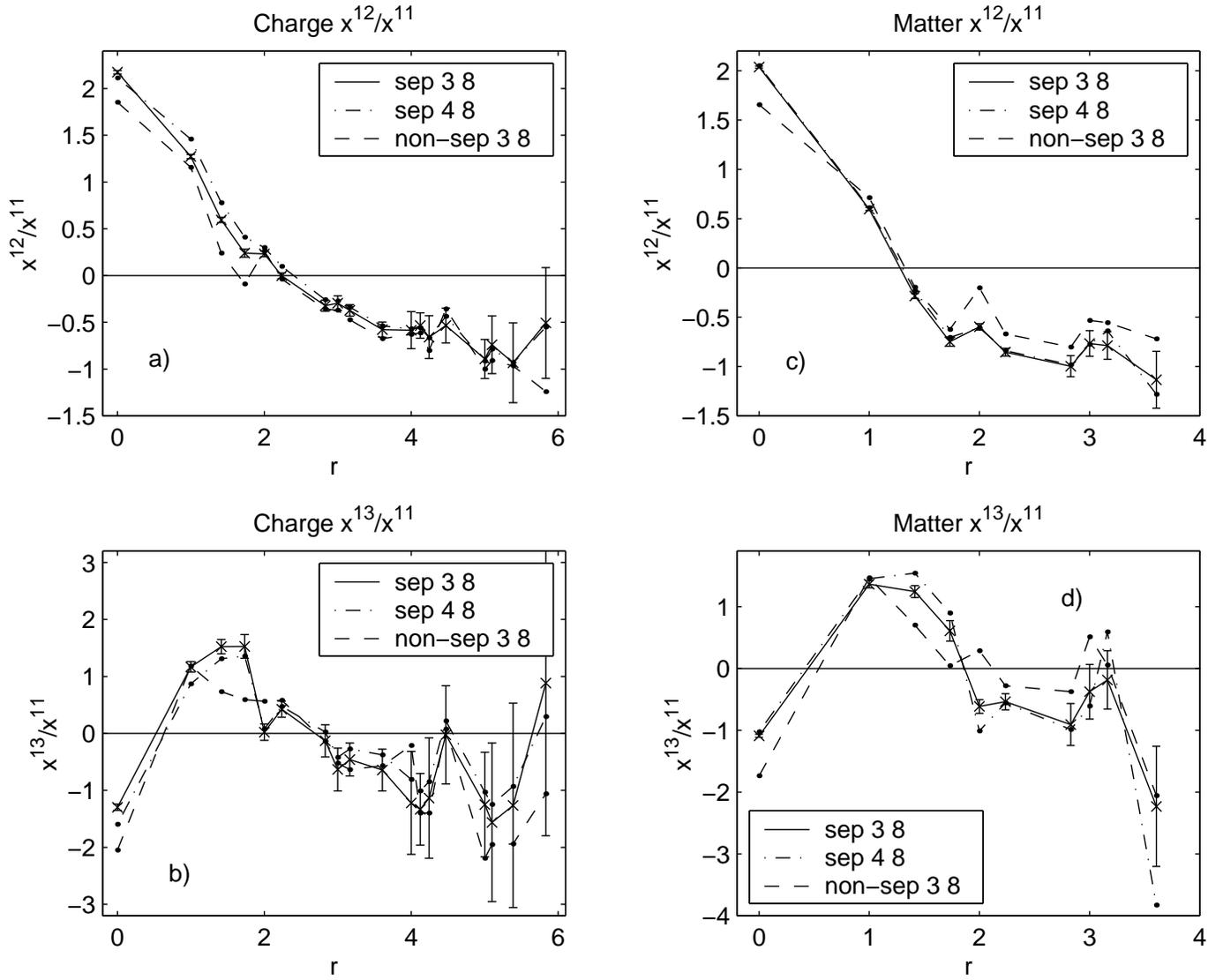} 
\caption{a) and b) --- the ratios $x^{12}/x^{11}$ and  $x^{13}/x^{11}$ 
for the charge distribution. c) and d) these ratios for the matter distribution}
\label{x12x11} 
\end{figure}

In Figure~\ref{x12x11}a), for  the charge density ratio $x^{12}/x^{11}$,
only the errors for the separable analysis with $T_{2,{\rm min}}=3$
and $T_{3,{\rm min}}=8$ are shown, since the other analyses have
similar errors. There a distinct node is seen at about 2.2 lattice
spacings i.e. at $\approx 0.3$~fm. Such a node is natural for $x^{12}$,
since it involves the excited S-wave state.
On the other hand, for the matter density --- as seen in
Figure~\ref{x12x11}c) --- the node is near to 1.5$a \approx 0.2$~fm.
 Figure~\ref{x12x11}b) shows the various analyses for the charge density
ratio $x^{13}/x^{11}$.
Here the node structure is less clear. One node is seen at about
2.8 lattice spacings i.e. at  $\approx 0.4$~fm. But a second possible
node at about 0.6 lattice spacings, i.e. at  $\approx 0.1$~fm, depends
on the one value of $x^{13}$ at $r=0$. However, we have no reason to
suspect that this is purely a lattice artefact. Furthermore, for second
excited S-wave states a second node is not unexpected.
Similar comments hold for the matter density ratio $x^{13}/x^{11}$ in
Figure~\ref{x12x11}d).
The node structure of  $x^{22}$ --- the charge density of the first excited
state --- is not at all clear. If $x^{22}$ is expressed in the separable
form 
$y_2(r)y_2(r)$, then the zero that should appear at about $2a$ is
not seen very distinctly in comparison 
with that seen in Figures~\ref{x12x11}a) or c) for $x^{12}/x^{11}$.

\begin{figure}[ht]
\includegraphics{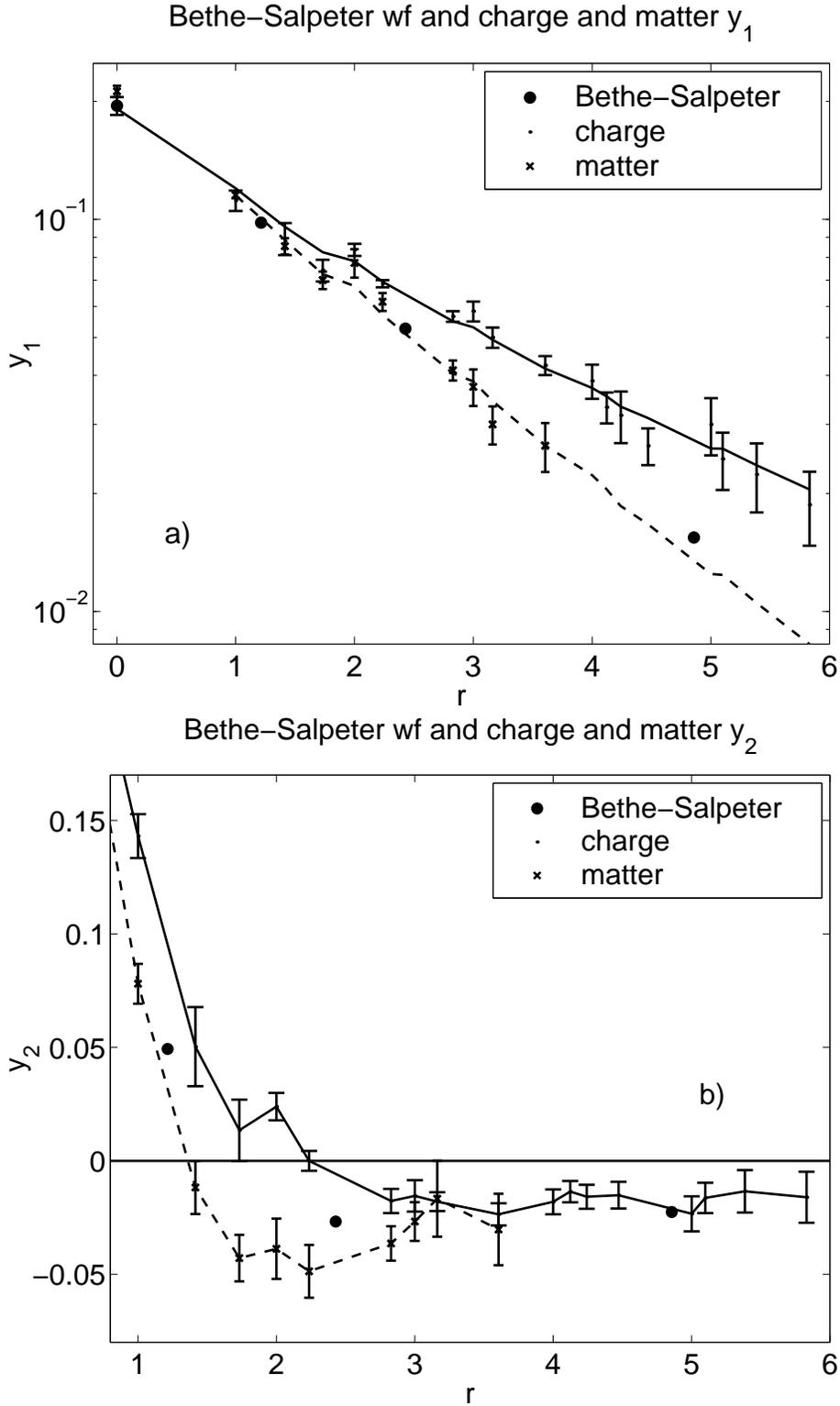}
\caption{The separable "wavefunctions" $y_1$ and $y_2$ defined in
the density $x^{\alpha \beta}(r)=y_{\alpha}(r) y_{\beta}(r)$.
\newline a) $y_1$ --- Solid line for the charge,  dashed for the matter,
   solid circles the Bethe-Salpeter wavefunction from
Ref.~\protect\cite{MP98},
\newline b) $y_2$ --- notation as in a) }
\label{y1y2}
\end{figure}

The above figures show directly the various charge densities 
$x^{\alpha \beta}(r)$. However, it is also of interest to see the
structure of the individual terms $y_{\alpha}(r)$ in the separable
form $x^{\alpha \beta}(r)=y_{\alpha}(r) y_{\beta}(r)$. These are shown
in Figure~\ref{y1y2} for both the charge and the matter. 
Figure~\ref{y1y2}a)  shows clearly that  $y_1(r)$ for the charge has a
significantly longer range than for the matter. Also as seen in 
Figure~\ref{y1y2}b) both of the  $y_2(r)$ exhibit a distinct node and are
responsible for the nodes in the separable form of the density  $x^{12}$
already seen in Figure~\ref{x12x11}a). We do not plot $y_3(r)$, since
the signal/error ratio is too small. 

As discussed at the end of the Introduction, the $y_{\alpha}(r)$ can
possibly 
be 
interpreted as wave functions for the state $\alpha$. However, there are
other radial distributions  associated with the $Q\bar{q}$ system that
can also be interpreted as wave functions. These are the Bethe-Salpeter
wavefunctions [$w_{\alpha}(r)$] discussed in Ref.~\cite{MP98}. 
They were extracted by assuming  the  hadronic operators 
$C_{2,\alpha \alpha}(r_1, \ r_2, \ T)$ to be of the
form $w_{\alpha}(r_1)w_{\alpha}(r_2)\exp(-m_{\alpha}T)$, where the sink
and source operators are of spatial size $r_1$ and  $r_2$. 
In Figure~\ref{y1y2} a comparison is made between
the above values of $y_1(r)$ and $y_2(r)$ and the corresponding results
form Ref.~\cite{MP98} for $w_1(r)$ and $w_2(r)$, where the latter have been 
normalised so that $w_1(0)=y_1(0)$ and $w_2(0)=y_2(0)$ and the values of
$r$ scaled.  Even though they  do bear some similarities, it
should be added that there are several reasons why these two types of wave
function should {\em not} agree in detail with each other. In particular, the 
$[w_{\alpha}(r)]^2$ cannot be identified as a charge or matter
distribution. In addition, they were found by using an explicit fuzzed path 
between $Q$ and $\bar{q}$ and so are dependent on the fuzzing
prescription, whereas the $y_{\alpha}$ are defined in a path-independent way.
Furthermore, in Ref.~\cite{MP98} the $w_{\alpha}(r)$ were extracted 
in the quenched approximation on a $12^3\times 24$ lattice
with the parameters $\beta=5.7$,
$C_{{\rm SW}}=1.57$ and $\kappa=0.13843$ --- the latter corresponding to a 
light-quark mass of  $m'_{\bar{q}}=1.77(4) m_{{\rm s}}$ 
i.e. about two strange quark masses.

\subsection{Charge sum rule}
\label{Chsr}
 In addition to measuring $C_3(r)$ for various values of $r$, the 
correlation where $r$ is summed
over the whole lattice is also obtained. This leads to
the charge sum rule as discussed in Ref.~\cite{G+K+P+M}.
The actual values of this sum rule are extracted using Eq.~\ref{C3fit},
where the $x^{\alpha \beta}$ are now independent of $r$.
The outcome --- as shown in Table~\ref{Chsumrule} --- is that $x^{11}$
is $\approx 1.3(1)$ consistent with the earlier quenched result.
The fact that $x^{11}$ is not unity --- as expected in the continuum 
limit --- can be qualitatively understood by introducing a 
renormalisation factor of $\approx 1/1.3 \approx 0.8$ into the basic $\gamma_4$
vertex used to measure the charge density. Such a factor of this
magnitude is reasonable as shown in Ref.~\cite{SRule}.

It is also reassuring that the $x^{\alpha \beta}$ with $\alpha \not= \beta$
are, in general, consistent with zero --- as expected in the continuum
limit. However, the interpretation of $x^{22}$ is less clear. The
non-separable (NS) case gives $x^{22}\approx 1.0(4)$ --- again a
reasonable value in the continuum limit --- whereas the separable case (S)
yields $x^{22}\approx 0.0$. This suggests that the separable
approximation may be less appropriate for excited states. 
 
In Table~\ref{Chsumrule} we also show the matter sum rule. These have a
somewhat wider spread of values with 0.9(1) being a reasonable
compromise --- a number that is about twice the estimate of 0.38(15)
for the quenched calculation of Ref.~\cite{G+K+P+M}. Perhaps this is an
indication --- unlike the matter radial distributions in 
Figure~\ref{x11vs2LY} --- 
that the quenched and unquenched results can differ even with
the present sea quark masses of about  the strange quark. 
However, we do not have the data to cross check with
 Refs.~\cite{foster} and \cite{Fosterthesis}, which
advocate the existence of such a difference for the matter sum rule.
   
\section{Form of radial dependence}
\label{radialdep}

The results in Table~\ref{Tablechden} are presented as simply a series
of numbers for each value of $r$ or $(x,y,z)$. However, it would be more
convenient and perhaps illuminating, if they could be parametrized
in some simple way. This can be done either in coordinate or
 momentum space --- the topics of the next subsections.  

\subsection{Coordinate space fit}
\label{CSF}

 Here we assume that the radial dependences of the lattice data in
Table~\ref{Tablechden} can be represented in terms of
exponential (E), Yukawa (Y) or gaussian (G) functions. 
Then the strategy is to first fit the
data at the largest values of $r$ with a single form in order to 
parametrize the longest ranged part of the data, where it is
expected that lattice artefacts will be less. This range is then 
 used as a starting point, when the data at all values of $r$ 
are fitted by adding a second form.

The reason for using exponential and Yukawa radial functions is that they
arise naturally as propagators in quantum field theory --- usually
in their momentum space form $(q^2+m^2)^{-1}$. However, if we go away
from  quantum field theory and attempt to understand the radial 
dependences in terms of wavefunctions from, for example, the Dirac
equation, then gaussian forms can then arise.

\subsubsection{Exponential, Yukawa and gaussian fits to the
charge and matter densities. }
\label{EYGden}
Assuming simply an exponential $a^{{\rm E}}\exp[-r/r^{{\rm E}}]$,
Yukawa $a^{{\rm Y}}\exp[-r/r^{{\rm Y}}]/r$ or gaussian
$a^{{\rm G}}\exp[-(r/r^{{\rm G}})^2]$ form and,
if all 18 pieces of  data are used (excluding that for 
$r=0$ in the Yukawa case), then the $\chi^2/n_{{\rm dof}}$ are  greater 
than unity --- being 1.7, 2.3 and 5.1 for E, Y, G respectively. 
However, these $\chi^2/n_{{\rm dof}}$ can be reduced significantly 
by adding a second exponential, Yukawa or gaussian form with the first
form having its amplitude $a_0^{{\rm F}}$ and range $r_0^{{\rm F}}$ 
guided by   those values needed to fit the data with $r\ge 3$. When these
combinations are used to fit all the data (excluding that for
$r=0$ in the Yukawa case) we get the results in the columns marked as 
2E, 2Y and 2G of Table~\ref{LFits}.  We see that the three radial integrals
$I^{{\rm F}}$ are comparable at $\approx 1.4(1)$ 
and  in reasonable agreement with those obtained earlier
by directly summing over the whole lattice and shown in Table~\ref{Chsumrule}.
These will be discussed later in Section~\ref{CSR},
where a detailed analysis of the charge sum rule will be made.

The various merits and defects of the E, Y and G forms will be discussed
in more detail later in the context of the separable approximation for
the density. At this stage all three forms are equally reasonable giving
$\chi^2/n_{{\rm dof}}$ that are $\approx 1.4$.

In Table~\ref{LFitsm} are shown the corresponding fits to the matter
density from column 5 of Table~\ref{Tablechdenex}. Here it
is seen that for 2E and 2Y the $\chi^2/n_{{\rm dof}}$ are
$\approx 1.7$ --- larger than in the charge case. In comparison,
for 2G the situation is reversed with a $\chi^2/n_{{\rm dof}} = 1.11$.

\subsubsection{Lattice exponential, Yukawa and gaussian fits to the
charge and matter densities. }
\label{LEYGden}
\begin{figure}[ht]
\includegraphics{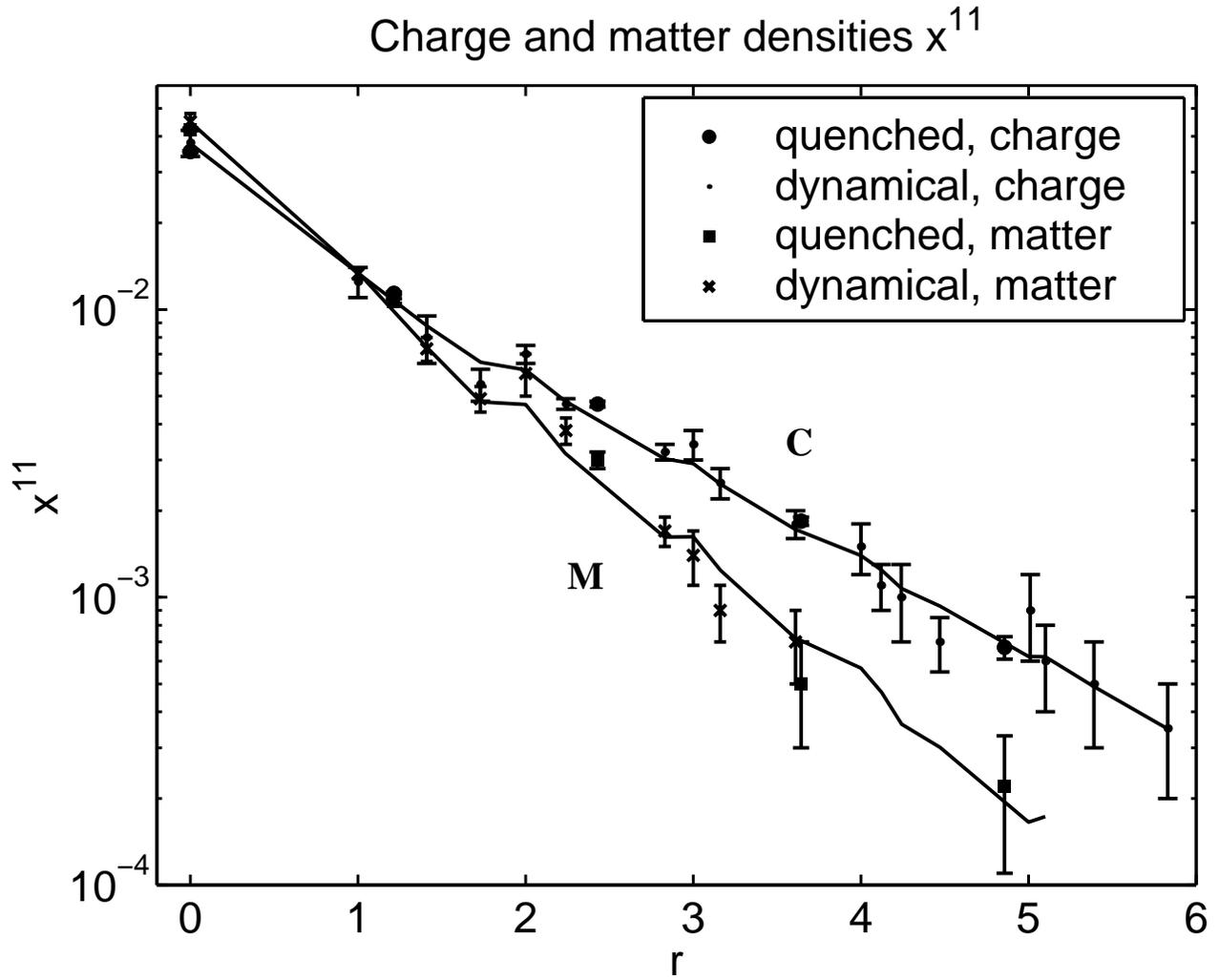}
\caption{The ground state charge (C) and matter (M) densities 
[$x^{11}(r)$] as a
function of $r/a$. The lines shows a fit to these   densities with a sum
of
two lattice exponential functions. The scaled quenched results of
Ref.~\protect\cite{G+K+P+M} are also shown by filled circles and
squares.}
\label{x11vs2LY}
\end{figure}

\begin{figure}[ht]
\includegraphics{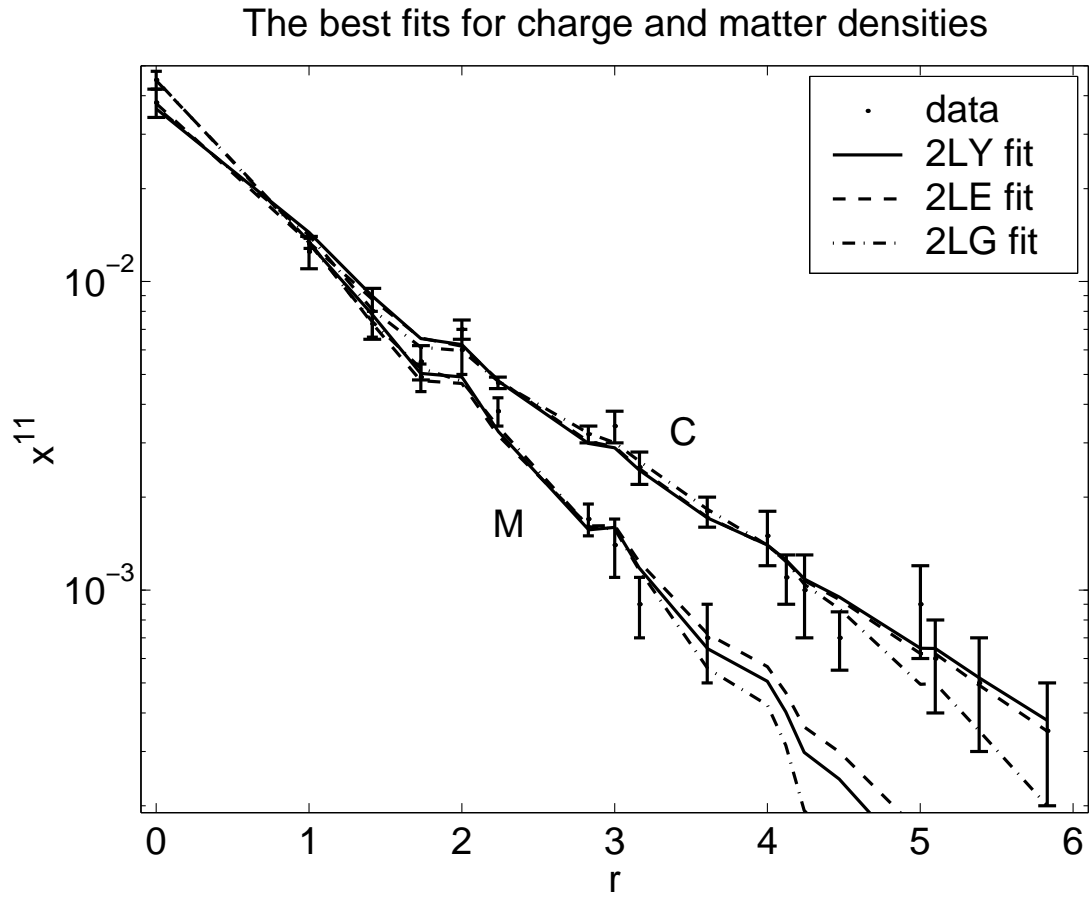}
\caption{Fit to the lattice data in Table~\protect\ref{Tablechden}
 with  lattice exponential (2LE), Yukawa (2LY) and gaussian (2LG)
forms in Table~\protect\ref{LFits}.} 
\label{LYEf} 
\end{figure}

As seen in Fig.~\ref{x11vs2LY}, the data is far from being
a smooth function of $r$ --- indicating lattice artefact effects.
Therefore, in addition to fitting the data with pure exponential, Yukawa
and gaussian forms, lattice versions of these (LE, LY, LG) are  also used.
In Ref.~\cite{L+R} the lattice form of the Coulomb function $(1/r)$ is
written as 
\begin{equation}
\label{LC}
\left[\frac{1}{{\bf r}}\right]_{{\rm LC}}=\frac{\pi}{aL^3}\sum_{{\bf q}}
\frac{\cos({\bf r}.{\bf q})}{D}.
\end{equation}
Here $L$ is the lattice size along one axis
 and $D=\sum^3_{i=1}\sin^2(aq_i /2)$, where
$ aq_i=0, \ \frac{2\pi}{L}, \ ... \frac{2\pi( L-1)}{L}, \ {\bf q}
\not=0$.
In this subsection, for clarity, the lattice spacing $a$ is shown explicitly.
For the above Yukawa form, Eq.~\ref{LC} is easily generalised to
\begin{equation}
\label{LY}
\left[\frac{\exp(- {\bf r}/r^{{\rm LY}})}{{\bf r}}\right]_{{\rm LY}}=
\frac{\pi}{aL^3}\sum_{{\bf q}}
\frac{\cos({\bf r}.{\bf q})}{D+0.25[a/r^{{\rm LY}}]^2}.
\end{equation}
However, now the point ${\bf q}=0$ can be included in the sum,
since it is no longer a singularity --- provided $1/r^{{\rm LY}} \not=0$. 
Similarly, lattice forms of the  exponential and gaussian can be obtained
by simply replacing in the usual Fourier transform the $q^2$ factors 
by their lattice equivalent 
$\frac{4}{a^2}\sum^3_{i=1}\sin^2(aq_i /2)=\frac{4D}{a^2}$ .
This results in
\begin{equation}
\label{LE} 
\left[\exp(- {\bf r}/r^{{\rm LE}})\right]_{{\rm LE}}=
\frac{\pi a}{2r^{{\rm LE}}L^3}\sum_{{\bf q}}
\frac{\cos({\bf r}.{\bf q})}{[D+0.25(a/r^{{\rm LE}})^2]^2},
\end{equation} 
\begin{equation}
\label{LG} 
\left[\exp[- ({\bf r}/r^{{\rm LG}})^2]\right]_{{\rm LG}}=
\left[\frac{r^{{\rm LG}}\sqrt{\pi}}{aL}\right]^3\sum_{{\bf q}}
\cos({\bf r}.{\bf q})\exp[-(r^{{\rm LG}}/a)^2D].
\end{equation} 
When using these
lattice forms,  columns 2LY, 2LE and 2LG in Tables~\ref{LFits} and 
\ref{LFitsm} for the charge and matter respectively show the  
results corresponding
to the usual forms 2Y, 2E and 2G.  There it is seen that the values of 
$r_0^{{\rm F}}$ for all
three  lattice  forms 2LE, 2LY and 2LG are quite similar to their 
2E, 2Y and 2G counterparts.
The outcome is that the lattice forms  are able to reproduce some 
of the structure 
especially near $r=2$ and 3 --- as seen in Figures~\ref{x11vs2LY} and \ref{LYEf}.
This is also reflected in the $\chi^2/n_{{\rm dof}}$'s being now less than unity
for the charge case and $< 1.2$ for the matter.
At first sight the two parametrizations 2LE and 2LY look 
very different, since 
 Eqs.~\ref{LY} and \ref{LE} have such dissimilar forms. In addition, in
 2LE the two terms add up, whereas in 2LY they  cancel,
since $a_0^{2{\rm LY}}$ and $a_1^{2{\rm LY}}$ have opposite signs in order to
dampen the $1/r$ effect at small $r$. 
However, later we shall see that, in practice, 2LE and 2LY behave
 in very similar ways --- with little numerical preference
for one over the other.

The conclusion from this subsection is that all three parametrizations
2LE, 2LY and 2LG are acceptable, since each can fit all the lattice data 
with $\chi^2/n_{{\rm dof}}$'s that comparable to unity.
\subsubsection{Exponential, Yukawa and gaussian fits to the
separable form of the charge and matter densities. }
\label{EYGSden}
The above has concentrated on fitting directly the ground state charge
density $x^{11}(r)$ in Eq.~\ref{C3fit}. However, in section~\ref{A3pf} a
second and possibly more natural parametrization --- a separable form $x^{\alpha
\beta}(r)=y_{\alpha}(r) y_{\beta}(r)$ --- was introduced. This resulted
in the data shown in  columns S of Table~\ref{Tablechden}. Here, we
consider that the $y_1(r)$ are simply $\sqrt{x^{11}(r)}$, where the latter are
the Best Estimate values in Table~\ref{Tablechden}.
The corresponding fits 2ES, 2YS and 2GS, with the above  forms  are shown in 
Tables~\ref{LFitswave} and \ref{LFitswavem}. For reasons to be seen
later, we only show the matter results for 2YS.
Several points can be seen:

\noindent 1) As expected, the ES and GS fits are now simply related to those
for the density  --- namely the $a_0^{{\rm F}}\approx (a_0^{{\rm FS}})^2$,
$r_0^{{\rm ES}}\approx 2r_0^{{\rm E}}$ and
$r_0^{{\rm GS}}\approx \sqrt{2}r_0^{{\rm G}}$.
Therefore, since little is added by the E and G fits, we do not show them
for the matter case.

\noindent 2)  When fitting all the data, the 2ES, 2YS and 2GS fits are 
comparable but all
with $\chi^2/n_{{\rm dof}}$'s that are greater than unity. When data points at small
$r$ are systematically removed, then the $\chi^2/n_{{\rm dof}}$ eventually
goes under unity. In the charge case for  2ES, 2YS and 2GS this happens at
$r\ge 2.24$, 2.24
and 2.0 respectively. However, the fact that the three $\chi^2/n_{{\rm dof}}$ are large
is partially due to the lattice artefacts already seen when dealing
directly with the density in the previous subsection.
 
\subsubsection{Lattice exponential, Yukawa and gaussian fits to the
separable form of the charge and matter densities. }
\label{LEYGSden}
 Here, the $y_1(r)$ are fitted  using the lattice exponential, Yukawa
and gaussian
forms in  Eqs.~\ref{LY}, \ref{LE} and \ref{LG} and the outcome is shown in 
Tables~\ref{LFitswave} and \ref{LFitswavem}.  
The fits 2LES, 2LYS, 2LGS  are slightly worse than
before --- but still acceptable.  In 2LYS the parameters
$a_0$ and $a_1$ are highly correlated with $a_1 \approx -a_0$ ---
again to dampen the $1/r$ effect for small $r$. In the limit
 $a_1 = -a_0$ for small $r$ the $1/r$ would cancel exactly to
leave a function more like an exponential. 

\vskip 0.5 cm

The conclusion to be drawn from this Section is that the lattice data 
extracted in Section~\ref{analysis} can be well fitted by any of the
three lattice forms in Eqs.~\ref{LY}, \ref{LE} or \ref{LG} ---
numerically none of them is superior and also none can be rejected.
This statement applies not only to the direct parametrizations of
the density as in Section~\ref{LEYGden} but also to the separable form
in  Section~\ref{LEYGSden}.

\subsection{Momentum space fit} 
\label{MSF} 
 
Often it is more convenient to view data in momentum space by making the
transformation
 \begin{equation} 
\label{FT}
x^{\alpha\beta}(\bf k)=
\sum_{\bf r} \cos(\bf k . \bf r)x^{\alpha\beta}(\bf r),
 \end{equation}
where the ${\bf r}$ summation should be over the whole 3-dimensional $L^3$
lattice. This would mean, in the present calculation, summing $x, \ y$ and $z$
over the ranges $-7$ to $+8$. If the lattice Yukawa, exponential and gaussian
expressions in Eqs.~\ref{LY}, \ref{LE} and \ref{LG}, written as
\begin{equation}
\label{LYE}
x({\bf r})=
\frac{\pi}{L^3}\sum_{{\bf q}}\cos({\bf r}.{\bf q})F({\bf q}),
\end{equation}
are now used to parametrize $x(\bf r)$, then we simply get
 \begin{equation} 
\label{FTYE}
x({\bf k})=\pi F({\bf k}).
\end{equation}
Here use has been made of the identity 
$\sum_{\bf r}\cos[{\bf r}.({\bf k}-{\bf q})]=L^3\delta_{{\bf k},{\bf q}}$.
From Eq.~\ref{FT} we see that the sum rule discussed earlier is
now directly $x({\bf k}=0)=\pi F(0)=I^{{\rm F}}$, where the
$I^{{\rm F}}$ are defined in the caption of Table~\ref{LFits}.  
The other fourier components with ${\bf k}\not=0$ are simply $\pi F({\bf
k})$.  

\subsubsection{The inclusion of lattice data directly into Eq.~\protect\ref{FT}}
The result in Eq.~\ref{FTYE} is not surprising, since it is
simply the Fourier transform of a Fourier transform. However, if ---
where available --- the lattice data of Table~\ref{Tablechden} are used
for the $x^{\alpha\beta}(\bf r)$  in Eq.~\ref{FT} and the form in 
Eq.~\ref{LYE} only used for the missing densities, then the result
should be an improved estimate of the $x^{\alpha\beta}(\bf k)$.
A measure of this can be obtained by fitting this improved estimate
with  the forms $F({\bf k},a_i,r_i)$ in Eqs.~\ref{LY}, \ref{LE} and
\ref{LG}, but where the parameters $a_i,r_i$ are tuned using {\sf Minuit} and are not
the ones appearing in Tables~\ref{LFits} and \ref{LFitswave}. 

As we
shall see in the next subsection, the fits of main interest are 2LY, 2LE
and 2LYS, since they possibly have a physical interpretation. However,
this  fitting procedure presents some problems, since we are attempting, 
with only four parameters ($a_{i=0,1},r_{i=0,1}$), to 
fit $165$ numbers --- 
the $x^{\alpha\beta}(\bf k)$ with $ k_i=0, \frac{2\pi}{L}, \ldots,
\frac{2\pi( L-1)}{L}$. 
In particular, if --- as a stability check ---
an analytic  form of $F({\bf k},a_i,r_i)$ is analysed using the same form,
then the {\sf Migrad} algorithm  of {\sf Minuit} does indeed give  {\em exactly} the correct
values of the $a_i,r_i$
but with large error bars. Another problem
is the choice of function to be minimized.  We consider two options:

(i) $N_1=[x_{\rm FT}({\bf k})-x_{\rm model}({\bf k})]^2$ and

(ii) $N_2=[x_{\rm FT}({\bf k})/x_{\rm model}({\bf k})-1]^2$, 

\noindent where
the $x_{\rm FT}(\bf k)$ are the Fourier transforms (with or without direct 
data inclusion) defined in Eq.~\ref{FT} and $x_{\rm model}(\bf k)$ are the
$\pi F(\bf k)$ in Eq.~\ref{LYE}. Since the $x_{\rm FT}(\bf k)$ decrease rapidly
as $\bf k$ increases, Option (i) emphasizes the smaller values of $\bf k$
and is appropriate for extracting  $r_0$, which is the longer range. In
principle, Option (ii) is better since it should give an overall fit to
the $x_{\rm FT}(\bf k)$. But in practice, it tends to be unstable yielding
either unacceptable solutions or very large error bars. 

If the 2LY fit to the charge density in Table~\ref{LFits} is analysed with a 
Yukawa form, then Options (i) and (ii) give  poles at $r_0=1.78(24)$
and $1.78(9)$ respectively --- see Table~\ref{polefit}. 
This is to be compared with the input value of $1.78(7)$ in Table~\ref{LFits}.
A similar strategy can be  applied to the improved 
$x^{11}(\bf k)$ generated from Eq.~\ref{FT}, where $x^{11}(\bf r)$
now contains directly lattice data wherever possible. 
In this case, Options (i) and (ii) yield $r_0=1.78(20)$ and $3.3(1.0)$ i.e.
the result from Option (i) is not distinguishable from using the
fitted expression in Eq.~\ref{LY} for all values of $r$, whereas that
from Option (ii) is unstable.
Similarly, if the 2LE fit in Table~\ref{LFits} is analysed with an
exponential form, then Options (i) and (ii) give  poles at $r_0=1.38(22)$
and $1.38(15)$ respectively. This is to be compared with the input value of
$1.37(6)$ in Table~\ref{LFits}. For the improved form of the density
the two Options yield $1.39(23)$ and $1.41(16)$. Again Option (i) is
indistinguishable from using only the fit values of $x^{11}(\bf r)$.
However, now option (ii) is more stable than before. 

In addition to analysing Yukawa forms with Yukawas or exponential forms
with exponentials other combinations are possible.
These are summarised, alongwith the above, in Table~\ref{polefit}.
Here the main interest is that, when analysing the exponential
forms with a Yukawa, the same value of $r_0=1.75(24)$ emerges.
 
The conclusion to be drawn from Table~\ref{polefit} is that the
parametrizations in Tables~\ref{LFits} and \ref{LFitswave} are so good
that any improvements on $r_0$ due to the inclusion of explicit lattice 
data cannot be detected.
Therefore, in the following discussions the Fourier transforms
based purely on Tables~\ref{LFits} and \ref{LFitswave} will, in general,
be used.

\subsection{Possible interpretations of the above fits.}
\label{inter}
In the above, the use of the various forms E, Y, \ldots, LYS,  LGS 
is considered as a purely
numerical exercise. However, one can also ask about any theoretical
interpretation of, or preference for, one form over the others. 
As we shall see below, in some cases,  this is best discussed in momentum space.
Also, since the inclusion of the original data has little effect, the fourier
transform of the charge density should be well described by simply
$\pi F({\bf k})$. Therefore, the discussion below focusses on the
interpretation of the different forms of $\pi F({\bf k})$. 
Here several possibilities are suggested:

\vskip 0.2 cm

1) {\bf Y and LY}. These forms can be  directly identified as the
propagators of a single particle. In the Y form ---
$a_i^{{\rm Y}}\exp[-r/r_i^{{\rm Y}}]/r$ --- the masses of
the propagating particles are simply
$1/r_i^{{\rm Y}}$. Also  for LY by writing the denominator
of Eqs.~\ref{LY} in the continuum limit as
$\frac{q^2}{4}+0.25[a/r_i^{{\rm LY}}]^2$, we see that these
masses can also be  identified as $1/r_i^{{\rm LY}}$. For
the charge case these will be vector particles, whereas for
the matter case they are scalars. From the fits to the charge
distributions in Table ~\ref{LFits}, the vector masses that
emerge are --- in lattice units of about 1.4~GeV ---
$am^{v}_{0,2{\rm LY}}=0.56\pm 0.02$ and $am^{v}_{1,2{\rm LY}}=0.89\pm 0.11$
with the corresponding scalar masses from the matter distribution in 
Table~\ref{LFitsm} being $am^{s}_{0,2{\rm LY}}=1.09\pm 0.06$ and 
$am^{s}_{1,2{\rm LY}}=1.4\pm 0.2$.

\vskip 0.2 cm
 
\begin{figure}[hb] 
\includegraphics{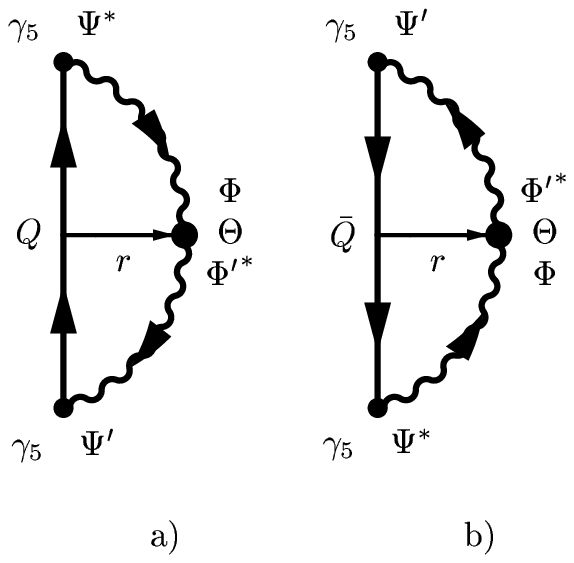}
\vspace{11cm} 
\caption{The two contributions to the three-point correlation function
$C_3$} 
\label{c3diags} 
\end{figure}

2) {\bf E and LE}. Looking at Fig.~\ref{c3diags} for
the three-point correlations, a cut in the $T$-direction intersects the two
light-quark 
propagators $G_q(t_1\rightarrow 0)$ and $G_q(0\rightarrow -t_2)$. 
In comparison, the lattice exponential form in Eq.~\ref{LE} also
contains a product of two propagator-like terms $1/[D+0.25(a/r^{{\rm E}})^2]$.
This  then suggests that the exponential form may be interpreted as the 
{\em product of two non-interacting quark propagators}.
Assuming  that the momentum of the probe is
divided equally between the two propagators, then the appropriate momentum 
transfer in each propagator becomes $q/2$. Now, when going to the continuum
limit, the denominators in Eqs.~\ref{LE} give the masses of the
propagating particles as $1/2r_i^{{\rm LE}}$ i.e. in the charge case 
$am_{0,2{\rm LE}}=0.36\pm 0.02$ and $am_{1,2{\rm LE}} \approx 4.5$. 
Therefore, one interpretation of
$m_{0,2{\rm LE}}$ is that this is the mass of a $\em constituent$
quark used in the naive quark model description of the meson
as simply two non-interacting  quarks. This would give a vector mass
 of $am^{v}_{0,2{\rm LE}}=0.72 \pm 0.04$.
A similar interpretation can be made for the matter case to give
a scalar meson of mass  $am^{s}_{0,2{\rm LE}}=1.07\pm 0.05$.

It should be added that in our earlier work in Ref.~\cite{G+K+P+M} using
the quenched approximation, the data were so sparse that an overall fit with
only a single exponential was attempted over a limited range of $r$
values --- the overall fit with a single Yukawa being much worse.
Therefore, to compare with the above values of $am^{v,s}$, a single Yukawa
fit to the data at the largest values of $r$ for which the data were still
reliable was carried out. For case 3 in Ref.~\cite{G+K+P+M}, the
charge density data at $r=3$ and 4 gave $am^{v}$=0.6(1) --- scaled to 
$a=0.14$ fm. Similarly, the matter density data at $r=2$ and 3 gave 
$am^{s}$=1.0(2).
These values  are not significantly different to the present estimates. 

In the above, the masses have been extracted by a somewhat tortuous
argument. However, in the literature there have been direct calculations
of the energies of these $q\bar{q}$ states using the
same lattice parameters and lattice size as those employed here. In 
Ref.~\cite{Allton} they got $am^{v}_0=0.785\pm 0.009$ 
and in Ref.~\cite{McN+M} $am^{s}_0=1.18\pm 0.08$.
These numbers are  consistent with  our above estimates from
the 2LE fit i.e. $0.72\pm 0.04$ and $1.07 \pm 0.05$ respectively. 
Also from the 2LY fit,  the scalar mass of $am^{s}_{0,2{\rm LY}}=1.09\pm 0.07$ is
consistent with this value. But the vector mass of 
$am^{v}_{0,2{\rm LY}}=0.56\pm 0.02$ is somewhat too small. 
This difference can be considered as a measure of the systematic error
on $m^{v}_0$ and suggests that the present data do not extend to
sufficiently large values of $r$ for a reliable estimate to be made of the
asymptotic form.

Unfortunately, it is not straightforward to identify the above
particles and their masses directly with physical particles, since we
use light quarks that have the isotopic spin properties of $u,d$-quarks
but with  masses about that of the strange quark. In addition, we do not
calculate  mass or density contributions that arise from
disconnected correlations.  For vector mesons, the latter has been shown in 
Ref.~\cite{Sharkey} to be only a small effect i.e. the OZI rule is
justified in this case. On the other hand, this appears  not to be so for scalar
mesons. Evenso, in the vector case it is not reasonable to identify the above
range of values  $m^{v}_0=0.9 \pm 0.1$ and $m^{v}_1=1.4 \pm 0.3$~GeV 
directly with
the isovector $\rho(0.77)$ and the radial excited $\rho(1.45)$ from 
Ref~\cite{PDG}, since in the quark model, the latter are
constructed from $u,d$ quarks with the correct mass --- a value much less than
the strange quark mass used here.
 However, since the OZI rule is a very good approximation 
for vector mesons, the additional mass of a state with strange quarks can be 
taken from the $\phi(1.02)$ meson --- a value somewhat larger than
our estimate of $m^{v}_0=0.9 \pm 0.1$~GeV. 
For the scalar mesons,  the comparison with experiment is even more
indirect. When comparing with our results,  from Ref.~\cite{PDG} 
the appropriate states
would be the  $a_0(0.98)$ and $a_0(1.45)$, since our neglect of disconnected
correlations effectively results in an isovector operator. However, as recently
discussed in Ref.~\cite{Close}, these states probably have a complicated
structure being mainly $(qq)_{\bar{3}}(\bar{q}\bar{q})_3$ in S-waves at 
short distances, with some $q\bar{q}$ in P-waves. But further out these 
rearrange into $(q\bar{q})_1(q\bar{q})_1$ and finally emerge as meson-meson
states.

In fact this complication arises also in lattice
calculations, since the present lattice parameters predict a
pseudo-scalar meson (the "pion") with mass $am^{\rm ps}=$0.564 --- see Table 8 in 
Ref.~\cite{Allton}. This means that in Fig.~\ref{c3diags} 
a cut  in the $T$-direction can intersect {\em four} light-quark
propagators --- a state that can be interpreted as the exchange of
two pseudo-scalar mesons. In the above scalar meson the $q\bar{q}$ are in a 
relative P-wave, so that it can couple to a two-meson state, where the
mesons are in a relative S-wave. On the lattice such a two pseudo-scalar
meson state would simply have a mass of $2am^{\rm ps}=$1.13. 
This means that the scalar meson --- calculated directly as a $q\bar{q}$
state with mass $am^{s}_0=1.18\pm 0.08$ in Ref.~\cite{McN+M} --- is
essentially degenerate with the two pseudo-scalar meson state 
and suggests that our
estimate of, say, $am^{s}_{0,2{\rm LY}}=1.09\pm 0.07$ also contains this
two-meson effect. It should be added that this problem does not arise
for the vector meson, since there the $q\bar{q}$ are in a relative
S-wave, so that the corresponding two-meson state has a relative P-wave
leading to  an energy considerably larger than $2am^{\rm ps}=$1.13. This
means that the structure of the vector meson generated here is expected
to be mainly $q\bar{q}$ with little mixing with the two-meson state.

\vskip 0.2 cm
 
 3) {\bf YS and LYS}.
 The above considers directly the charge density. 
However, a similar analysis  can be carried
out on the $y_1(r)$, when the charge density is expressed in the separable form
 $x^{11}(r)=y_{1}(r) y_{1}(r)$.
Since the single form fits ES, GS, LES and LGS  are given  directly by their
non-separable counterparts, as is clear by the relationships
$a_0^{{\rm F}}\approx (a_0^{{\rm FS}})^2$,
$r_0^{{\rm ES}}\approx 2r_0^{{\rm E}}$ and
$r_0^{{\rm GS}}\approx \sqrt{2}r_0^{{\rm G}}$, 
they  would add nothing new. The same is essentially true of the 
2ES, 2GS, 2LES and 2LGS fits in Table~\ref{LFitswave}, since the
values of the $r_0^{{\rm FS/LFS}}$ and $r_0^{2{\rm FS}/2{\rm LFS}}$
are very similar. However, if the $y_1(r)$ are thought of  as
one of the two propagators in Figure~\ref{c3diags}, then it is now
appropriate to identify $y_1(r)$ with the YS and LYS forms.
From Table~\ref{LFitswave} we get that  $m_{0,2{\rm LYS}}=0.49\pm 0.02$
and $m_{1,2{\rm LYS}}=0.58\pm 0.02$~GeV.

This value of $m_{0,2{\rm LYS}}$ is consistent with the earlier
$m_{0,2{\rm LE}}=0.51\pm 0.03$~GeV and supports the identification that
these two very different ways of analysing the data are indeed extracting
the propagator of the same "particle" and, as said above, it is tempting
to identify this  "particle" with a constituent  quark. 
Since the density in the 2LE case can be expressed schematically in terms of
two Yukawas ($Y_i$) as $Y_0Y_0+Y_1Y_1$ compared with the corresponding  
expression in the 2LYS case $(Y'_0+Y'_1)^2$, it is really only
justified to compare  the dominant terms $Y_0Y_0$ and $(Y'_0)^2$.

In Table~\ref{LFitswavem} we show the corresponding fits to the matter 
distribution giving $m_{0,2{\rm LYS}}=0.63\pm 0.05$ and
$m_{1,2{\rm LYS}}=1.0\pm 0.2$~GeV. {\em If} these are now interpreted
as the masses of  constituent quarks, then scalar mesons constructed
from two such {\em non-interacting} quarks --- as in the naive quark
model ---  would predict $m^{s}_0=1.26\pm 0.10$ and $m^{s}_1=2.0 \pm 0.4$~GeV.
The latter is consistent with $m^{s}_1(2{\rm LY})=2.0 \pm 0.3$~GeV
discussed earlier. However, the value of $m^{s}_0$ is distinctly smaller
than the earlier $m^{s}_0(2{\rm LY})=1.53 \pm 0.09$~GeV and supports
the point of view that scalar mesons cannot be described by the naive
constituent quark model.
\vskip 0.2 cm
 
4) {\bf GS and LGS}. 
The above Yukawa and exponential forms arise naturally in quantum
field theory, whereas gaussians do not.
However,  when --- as in Ref.~\cite{G+K+P+M} ---  
an attempt is made to
understand the densities in terms of  solutions of the Dirac equation,
it would be difficult to reconcile an exponential or Yukawa asymptotic
tail with the usual form of linearly rising confining potential $cr$. When
such a potential is introduced as a  scalar  potential, the solutions of
the Dirac
equation are asymptotically gaussian. This is most easily seen when --- 
in the notation of Ref.~\cite{B+D} --- the coupled 
Dirac equation for large $r$ is written as
\begin{equation}
-m(r)G(r)=-F'(r) \ \ {\rm and} \ \ m(r)F(r)=G'(r) \ \  {\rm giving} \ \ 
G''+(cr)^2G=0,
\label{Dirac}
 \end{equation}
where $m(r)=m+cr\rightarrow cr$. 
The functions $G$ and $F$ are, therefore, seen to
decay asymptotically as  gaussians. Of course, the concept of a linearly 
rising confining potential does not hold for sufficiently large $r$,
since eventually this will be quenched by the creation of 
$q\bar{q}$ pairs. Unfortunately, the actual demonstration of this
unavoidable effect has
yet to be achieved in a completely convincing manner for full QCD.
However, the indications are that this will only occur for some value of
$r$ greater than about 1.2 fm, which --- with the present lattice spacing of 
$a \approx 0.14$~fm --- corresponds to a distance of almost $10a$.
Therefore, in the range of interest here ($4a < r < 7a$) the linearly 
rising  potential is still expected to be important and its repulsion
could well suppress the density from  being an exponential decay to more
like a gaussian decay. 

Support for a Dirac equation description is also given by our result
that the charge and matter distributions are different --- a feature not
easy to understand in a non-relativistic approach.
In the notation
of Eq.~\ref{Dirac} the charge and matter distributions can be expressed as
$x_c^{\alpha\beta}(r)=G_{\alpha}(r)G_{\beta}(r)+F_{\alpha}(r)F_{\beta}(r)$
and $x_m^{\alpha\beta}(r)=G_{\alpha}(r)G_{\beta}(r)-F_{\alpha}(r)F_{\beta}(r)$
respectively. Attempts are now underway to study to what extent the above 
 distributions can indeed be interpreted in terms
of solutions $(G,F)$ of the Dirac equation.

Our use, in section \ref{A3pf}, of the separable approximation when 
analysing  the data  was suggested by the form of the three-point 
correlation function $C_3$ and supported by the {\em presence in the
data} of the symmetry $R(T)=R(T+1)$ when $T$ is odd --- see Eq.~\ref{sym}.
However, as seen in Table~\ref{symt} this support is mainly
for small values of $r$ and is not present for the larger values of $r$.
The above interpretation of the data in terms of Dirac wavefunctions is
consistent with this --- namely --- for small $r$ the lattice data
gives $x_c^{11}\approx x_m^{11}$. This implies that $G_1(r) \gg F_1(r)$ and
both densities are approximately described by $G_1(r)G_1(r)$ --- a
separable form. Whereas at the largest values of $r$ studied here, we get 
$x_c^{11}\approx 3 x_m^{11}$ --- implying that $G^2_1(r)\approx
2F^2_1(r)$ and so destroying the simple separability of the densities.

\subsection{More on the charge sum rule.} 
\label{CSR} 

Comparing the values of the sum rule in Table~\ref{Chsumrule}, where the
estimates are  made by directly summing over all the lattice as in
Section~\ref{Chsr}, and the values of $I^{{\rm F}}$ in 
Table~\ref{LFits}, where they are made by summing the separate
contributions from each vertex $(x, \ y, \ z)$, we see that the latter
in the 2LY case can be as large as $1.7(3)$  --- a number that seems to be 
 slightly larger than the direct sum of about $1.4(1)$. 
Of course, within
the quoted error bars these two estimates agree. Even so there are reasons
 --- to be discussed below --- why exact agreement is not necessary. 
Firstly, it must be remembered
that the direct estimate includes contributions from the whole lattice
i.e. upto values of $r=8\sqrt{3}$, whereas the data for a given $r$ is
based on  only 18 values extending upto almost $r=6$. More explicitly,  
the fits using Eqs.~\ref{LY} and \ref{LE}
are to data with $r<6$ --- the largest being $x=5, \ y=3, \ z=0$. In
fact, a more correct statement would be that the fits are mainly dictated by
the data for $r\le 4$, with the remaining data, which have relatively
 large error bars, playing more of a supportive than a decisive role. 
Therefore, there is no guarantee that the expressions based on
 Eqs.~\ref{LY} and \ref{LE} are a good representation beyond $r\approx 4$.

To test the importance of the density contributions from large values of
$r$, the  summation over ${\bf r}$ in Eq.~\ref{FT} is 
carried out explicitly but truncated in two ways
by introducing either a cubic cut-off (C) or a spherical cut-off (S). In
the cubic case, the $x, \ y$ and $z$ sums are each limited to the  
range $-r_c \ \  {\rm to} \ \  +r_c$, where $r_c$ takes on  values ranging from 3
to 7. The first
 values of $r_c$ cover much of the range over which 
Eqs.~\ref{LY} and \ref{LE} were used to fit the data, whereas by going to
$r_c=7$  additional points are included  that are outside this range.
In the spherical cut-off, only values of $x, \ y$ and $z$ with
$x^2+y^2+z^2\le r_c^2$ are kept in the summation.
The outcome as seen in Table~\ref{cutoff} shows several points:

1) There is little difference between the use of two lattice 
exponentials (2LE) versus two lattice Yukawas (2LY). 
This  shows that, for the sum rule, the two forms are not only very similar
at those values of $(x, \ y, \ z)$ for which there exists lattice data
but also at all other points on the lattice.
Of course, when the expressions in Eqs.~\ref{LY} and \ref{LE} are used
over the whole lattice, there is no need to resort to the explicit summation
in Eq.~\ref{LYE}, since the result (Eq.~\ref{FTYE}) is known. However,
it did serve as a numerical check.

2) The effect of including --- where ever possible from 
Table~\ref{Tablechden} --- the 18 actual lattice
data points  instead of the fitted forms has the minor effect of
decreasing the sums by about $0.02$. This is yet another reflection that
the fitted forms are a good representation of the lattice data.
Of course, these 18 points are only a small fraction of the total
data needed in Eq.~\ref{FT}. Fortunately, about one half of the sum rule
comes from contributions within a volume where $r\le 3$ and these have
all been measured directly.  

3) The effect of using the Cubic (C)- versus the Spherical (S)-cutoff is large,
with the latter being consistently about $0.2$ smaller. This is natural, since
for the same value of $r_c$ the C-cutoff embraces more lattice points.

4) However, the most significant point is that the sums continue to 
increase significantly as $r_c$ goes beyond the
range where the lattice data is measured --- with the value at $r_c=5$
for the S(C)-cutoff being about  1.1(1.4) increasing to 1.4(1.6)
at $r_c=7$.

The outcome is that, from the region $r\le 4$, where the fits in
Eqs.~\ref{LY} and ~\ref{LE} are most reliable, 
 the contribution to the sum rule is about 1.2 -- 1.4, 
which already is consistent with the direct summation estimate in 
Table~\ref{Chsumrule}. This, therefore, means that there is a significant
contribution of about 0.4 -- 0.5 coming from values of $r$ greater than 4. 
In detail, a contribution of about 0.2 comes from the range
$4<r<5$ and about  0.1 from the range $5<r<6$, where the fits seem
to be supported by the data. The remaining discrepancy of almost 0.2
then comes from the periphery with $r\ge 6$, where there is no data to
check the meaningfulness of the fits. 

   It, therefore, seems that the direct sum rule in Table~\ref{Chsumrule}
could well be slightly smaller that the explicit sum in Eq.~\ref{FT}. 
This possible difference can be interpreted in two ways: The explicit sum is an
overestimate or the direct sum is an underestimate.

1) The fits using the expressions in 
Eqs.~\ref{LY}  and \ref{LE} may indeed give a good 
fit to the data upto $r\approx 4$, but are  {\em not}  good
estimates of the poor and missing data for $r>4$. To test this, the
lattice forms 2LE and 2LY  are used upto $r=r_c$ and the single non-lattice
forms based on 2E, 2Y and 2G in Table~\ref{LFits} are used for $r>r_c$. 
The outcome is shown in Table~\ref{Tails}. 
As expected the combinations 2LE+E and
2LY+Y give results in the range 1.4--1.5, which are   close to those of 
the spherical cutoff with $r_c=7$ 
in Table~\ref{cutoff}. On the other hand, the
combinations 2LE+G and 2LY+G --- containing the gaussian tail --- give a result
 of $\approx 1.3$, which is noticeably smaller than the 2LE+E or 2LY+Y 
numbers. Furthermore, all of these combinations result in sumrules
that are smaller than the $r_c=7$ {\em cubic} cutoff values of almost 1.7.

2) The second --- and less likely --- interpretation is that the fits with
Yukawa and exponential
forms do indeed give a good estimate of the poor and missing data for
$r>4$. This would mean that the values of the sum rules in 
Table~\ref{Chsumrule} are an underestimate and that the measured contributions
there from $r>4$ are too small. An example of this   will be seen
later in Table~\ref{rotv2} of the next subsection. There the 
three-point correlation function for $r=5$ is consistent with that
for $(x=3, \ y=4)$ for $T\le 6$ --- but with errors that are twice as big.
However, for $T>6$ --- the $T$ range necessary for extracting densities
 --- the signals from the two cases differ greatly, 
with those for $r=5$ simply disappearing.  Possibly similar
underestimates could occur for larger values of $r$,  where the fits from
Eqs.~\ref{LY}  and \ref{LE} suggest a  significant 
contribution ($\approx 0.3$) to
the sumrule should arise. However, this would be surprising since
it is not usual for there to be such systematic trends. Normally in such
cases, one would expect the results to fluctuate from being too low  for some
some values of $r$ to being too high at others.
However, if the trend suggested by the $r=5$ data were true 
then this would mean that the
charge sumrule could be larger than that measured directly --- possibly upto 
about 1.7. Unfortunately, to now get the
value of unity expected in the continuum limit, would then require
a vertex renormalisation factor of about 1/1.7 $\approx$ 0.6 --- a 
number significantly smaller than the estimates in Ref.~\cite{SRule}.

The first interpretation has two nice features. Firstly, with a charge
sumrule of about
1.3, the required vertex renormalisation factor to ensure unity in the 
continuum would be about 0.8 --- a value more in line with the estimates
in Ref.~\cite{SRule}. Secondly,  when --- as in Ref.~\cite{G+K+P+M} and
 discussed above ---  an attempt is made to
understand the charge density in terms of  solutions of the Dirac equation,
exponential and Yukawa forms do not arise naturally, whereas gaussian
forms do as illustrated in Eq.~\ref{Dirac}.

For the matter sum rule we saw in Table~\ref{Chsumrule} that $0.9(1)$ was a
reasonable compromise. The predictions using the algebraic fits are shown
as $I_{{\rm F}}$ in Table~\ref{LFitsm} and are seen to have much larger error
bars than their charge counterparts. All that can be said is that the
$I^{{\rm F}}$'s are --- within these large uncertainties --- consistent
with the direct measurement of 0.9(1).

The conclusion from this section is not definite. If the asymptotic form
of the density is indeed exponential or Yukawa, then the charge sumrule 
is 1.4 to 1.5. However, it is not possible to rule out an asymptotic
form that is  gaussian. In that case the  charge sumrule could be less
than 1.3.

\subsection{Rotational invariance}
\label{RotationI}

Since off-axis points are considered, in principle it should be possible
to check rotational invariance. In particular, comparisons between the
on-axis point $r=5$ and the off-axis point $(x,y)=(3,4)$ data and also the
$r=3$ and $(x,y,z)=(2,2,1)$
data are of special interest. 
In Table~\ref{rotv2} the ratios 
$\langle C_{3,{\rm F}_1{\rm F}_1}(T, \ r)\rangle /\langle
C_{2,{\rm F}_1{\rm F}_1}(T)\rangle $  are shown
for the dominant charge correlation. This table shows several points:

1) The $r=3$ and $(2,2,1)$ data both have good signals for all
values of $T$. This shows clearly that rotational invariance is violated
with $x_{{\rm F}_1{\rm F}_1}^{11}(r=3)$ being almost twice
$x_{{\rm F}_1{\rm F}_1}^{11}(2,2,1)$.
The reason for this lack of rotational invariance in such an extreme
off-axis case could be due to the
presence of the single $z=1$ step in $(2,2,1)$. A similar effect can be seen in 
Figure~\ref{x11vs2LY} for 
the $(1,1,1)$ data which is also lower than the general trend.
For less extreme cases such as $(2,1)$, rotational invariance seems to be 
better satisfied.

2) The $r=5$ signal disappears at $T=7$ - rendering it useless to
extract the density, since this requires reliable data for $T>7$.
This negative result is anyhow of interest, because in Ref.~\cite{G+K+P+M}
$r=5$ was the largest value of $r$ analysed and it suggested that
the charge density was considerably smaller than would be expected from 
a simple exponential dependence. This is no longer the case.

3) The $(x, y)=(3, 4)$ signal is good for all $T$. This improvement
over the $r=5$ case is presumably due to the fact that each off-axis
correlation is measured 24 times compared with the 6 on-axis
measurements.

4) For $T< 7$, the two sets of data are in agreement, as would be expected
if rotational invariance had been achieved. In fact, this comparison
can be pushed further by noting, that at these 3 values of $T$,
$C_3(r=5)$ appears to be slightly larger than
$C_{3,{\rm F}_1{\rm F}_1}(3, 4)$ --- as
would be suggested by the strong coupling model.

If the lattice Yukawa and exponential forms in section~\ref{radialdep}
are used to estimate the rotational invariance, we get the results 
in Table~\ref{rotvlye}. 
Both forms give the same results well within  $1\%$.
However, for the two cases considered involving directly the
lattice data --- $r=3$ versus
$(2,2,1)$ and $r=5$ versus $(3,4)$ --- the  
off-axis values are about $10\%$ less than the corresponding on-axis
value. For $r=3$ versus $(2,2,1)$ this difference is considerably smaller
than the actual lattice data requires. On the other hand,
for $r=5$ versus $(3,4)$ the reverse could be true with the difference
being much larger than is
suggested by the small $T$ values in Table~\ref{rotv2}.

It should be added that this study of the $(2,2,1)$ case was only carried
out after the bulk of this work was completed. In particular, the
(2,2,1) data was {\em not} included in the fitting that led to the
exponential, Yukawa and Gaussian parameters in Tables~\ref{LFits} to
\ref{LFitswavem}. However, its direct inclusion did not result in
values of the $a_i$ and $r_i$ that were outside of the error bars given
in Tables~\ref{LFits} to \ref{LFitswavem}, eventhough each
$\chi^2/n_{{\rm dof}}$ did increase by about 0.3 -- 0.4. A more realistic
way to treat this data is to combine the $r=3$ data  with that for
(2,2,1) using the relative weights of 6 to 24 --- the number of equivalent
possibilities for the two cases. When this weighting is also used for
the fitting functions, the $\chi^2/n_{{\rm dof}}$ decreases slightly from the
values in Tables~\ref{LFits} to \ref{LFitswavem}. However, again the
final values of the $a_i$ and $r_i$ are unchanged within the errors
quoted. This suggests that the overall fits are sufficiently stable 
that the effect of any particular case may lead to  a large 
$\chi^2$ contribution for that case but still leave the  fit 
essentially unchanged. It should be noted that the $r=3$ and $(2,2,1)$
charge data  only differ by about two standard deviations, so that the
above problem could even evaporate if more gauge configurations were
used.

\section{Conclusion} 
\label{con} 
In this paper charge (vector) and matter (scalar) radial distributions
have been measured on a lattice for the heavy-light meson ($Q\bar{q}$),
where $Q$ is a static quark and  $\bar{q}$ has a mass approximately that
of the strange quark. The charge distribution could be determined
reasonably well upto an interquark distance of about 6 lattice spacings
i.e. $\approx 0.8$~fm. In comparison, the matter distribution
measurements could only be carried out upto about 4 lattice spacings 
i.e. $\approx 0.6$~fm. The drop-off of the charge distribution
can be well described by the exchange of a vector meson
of mass $\approx 1$~GeV. 
On the other hand, the drop-off of the matter distribution is
described by the exchange of an scalar meson
of mass $\approx 1.5$~GeV. 

In the conclusion of Ref.~\cite{G+K+P+M} several refinements and
extensions to that pilot calculation of charge and matter distributions 
were listed. Here we have carried out a few of these:

1) Probably the most important "refinement" is the replacement of the
quenched approximation by the use of dynamical fermions. However,
as seen in Figure~\ref{x11vs2LY}, we find that the two appear to be
indistinguishable within the accuracy of the present work.
In Refs.~\cite{foster} and \cite{Fosterthesis} it was suggested that in
the matter distribution there could be a difference due to the presence
of disconnected quark-loop contributions. The fact that this is not seen
here could be due to our use of sea quarks that have a mass about that 
of the strange quark mass.

2) Radial correlations at off-axis points are now measured. This meant that
the number of data points accessible before the noise takes over is 
much larger --- going  from about 6 to 18. This enables us to
achieve better algebraic fits to the data. However, one of our hopes, to
see rotational invariance by comparing the data for $r=5$ and $(3,4)$
was only partially successful --- see Table~\ref{rotv2}.

3) The lattice spacing is now smaller --- 0.14~fm compared with
the earlier 0.17~fm. Also the number of gauge configurations is larger
--- 78 versus 20 earlier.

However, the list in Ref.~\cite{G+K+P+M} contained other points not
touched here:
 
1) So far we have only extracted $S$-wave correlations. 
We still need to  measure the  $P_{1/2}, \   
P_{3/2}, \ D_{3/2}, \   D_{5/2},...$ densities corresponding to the  
energies extracted in Ref.~\cite{MP98}. Also for a given orbital angular 
momentum, do these correlations show the  degeneracy predicted in    
Ref.~\cite{Page}? 

2) The measurement of correlations in the baryonic and
$(Q^2\bar{q}^2)$ systems. Are these similar to those in the $(Q\bar{q})$
system -- as is the case  when comparing correlations in 
few-{\em nucleon} systems? If this is so, then it would encourage 
phenomenological approaches such as that mentioned above utilizing the
Dirac equation.

3) As with all lattice calculations, there is the need to check the
continuum limit by  using finer and larger lattices. The former has, to
some extent, been checked by the comparison between our
earlier work in Ref.~\cite{G+K+P+M} with $a=0.17$~fm and this
study with $a=0.14$~fm. There, as seen in Figure~\ref{x11vs2LY},
the two sets of results agree after scaling. However, the two
calculations also differ by their use of  quenched versus
unquenched gauge configurations. A more correct comparison would
involve  the use of the same type of configurations --- quenched or 
unquenched. Even so, our present comparison --- in spite of its
failings --- is encouraging.

4) So far we have dealt with light quarks (valence and sea) that have 
the isospin form of $u,d$-quarks but with a
mass about that of the strange quark and heavy quarks that are static.  
The use of such  quark masses means that the nearest physical meson 
with which we can possibly compare our calculations is the $B_{{\rm s}}$(5.37~GeV).
However, for a more realistic comparison we should eventually have 
a heavy quark with mass about 5~GeV and $u,d$-quarks with their correct
mass.

5) In this work we have only probed the charge and matter distributions using 
 the $\gamma_4$ and unity probes. However, other probes are possible
such as: i) The pseudo-vector operator $(\gamma _{\mu}\gamma_5)$  needed
for the $B^*B\pi$ coupling --- see Ref.~\cite{Divitiis}; ii) Probes to
study the color structure of the gluon fields and possible $q\bar{q}$
condensates surrounding the $Q\bar{q}$ system.

This study is now at the stage where we have the energies and
the corresponding charge and matter radial distributions for the
ground and  first excited S-wave states. In the future we hope to
have these quantities for the other partial waves in Ref.~\cite{MP98}.
These results can possibly be utilized in at least two ways:

1) When, for example, calculating electromagnetic transitions between
different $Q\bar{q}$ states the form of the transition matrix elements
could be guided by the above radial distributions.

2) As discussed in the Introduction --- and also as partial motivation for
the separable method for analysing the lattice data --- the radial
distributions could be interpreted in terms of wavefunctions. 
This would mean we are in the position of having some of the eigenvalues and 
eigenfunctions of an Hamiltonian, whose form we do not know. We are then
at liberty to find this Hamiltonian to construct an effective theory. This
effective Hamiltonian could be, for example, of Schroedinger or Dirac form with
suitable interactions. As discussed in Section~\ref{inter}, the latter form
is probably more appropriate,
since the charge and matter distributions are different --- a feature not
easy to understand in a non-relativistic approach. 

This study has shown that --- using lattice techniques --- reliable estimates
can be made not only of spectra but also of wavefunction information.
This is just the beginning --- with future studies expected to enlarge this
information and also to attempt interpretations outside of quantum field theory.

\section{Acknowledgements}
The authors wish to thank the Center for Scientific Computing in Espoo, 
Finland for making available resources without which this project could 
not have been carried out. One of the authors (J.K.) wishes to thank the
Magnus Ehrnrooth Foundation for financial support.
Useful conversations with Nils T\"{o}rnqvist
are also acknowledged.
 
\appendix 
\section{The symmetry $R(T)=R(T+1)$ for odd $T$} 
\label{Asym}
In the separable case where $x^{\alpha \beta}(r)=y_{\alpha}(r) y_{\beta}(r)$,
for even $T$
\begin{eqnarray}
R(T,r)=&-\ln\Bigl[1+\frac{c(2,1) y_2(r)}{c(1,1)
y_1(r)}e^{-(m_2-m_1)\frac{T}{2}}+\frac{c(3,1) y_3(r)}{c(1,1)
y_1(r)}e^{-(m_3-m_1)\frac{T}{2}}\Bigr]\\
&+\ln\Bigl[1+\frac{c(2,1) y_2(r)}{c(1,1)
y_1(r)}e^{-(m_2-m_1)\frac{T-2}{2}}+\frac{c(3,1) y_3(r)}{c(1,1)
y_1(r)}e^{-(m_3-m_1)\frac{T-2}{2}}\Bigr],
\end{eqnarray}
whereas for odd $T$
\begin{eqnarray}
R(T,r)=&-\ln\Bigl[1+\frac{c(2,1) y_2(r)}{c(1,1)
y_1(r)}e^{-(m_2-m_1)\frac{T+1}{2}}+\frac{c(3,1) y_3(r)}{c(1,1)
y_1(r)}e^{-(m_3-m_1)\frac{T+1}{2}}\Bigr]\\
&+\ln\Bigl[1+\frac{c(2,1) y_2(r)}{c(1,1)
y_1(r)}e^{-(m_2-m_1)\frac{T-1}{2}}+\frac{c(3,1) y_3(r)}{c(1,1)
y_1(r)}e^{-(m_3-m_1)\frac{T-1}{2}}\Bigr].
\end{eqnarray}
Here we have dropped the fuzzing indices on $R$.
By inspection it is seen $R(T)=R(T+1)$ as given in Eq.~\ref{sym}
and demonstrated in Table~\ref{symt}.

\newpage 

\begin{table} 
\caption{Values of the parameters $am_{\alpha}$ and $ v_i^{\alpha}$,
 needed for fitting the $C_2$-correlations. 
\newline Case Q:  20 quenched configurations with $M_2=3$ and $T_{2,{\rm min}}=4$  
--- referred to as Case 3 in \protect\cite{G+K+P+M} 
\newline Case A: 78 dynamical configurations with $M_2=3$ and $T_{2,{\rm min}}=4$
\newline Case B: 78 dynamical configurations with $M_2=4$ and $T_{2,{\rm min}}=3$
\newline Case C: 78 dynamical configurations with $M_2=4$ and
$T_{2,{\rm min}}=4$
\newline The entries \{...\} show the values when correlations are
removed as discussed in the text.
\newline The entries [..] for $\Delta m_{21}$ and $\Delta m_{31}$ include the
ratio of lattice spacings 0.14/0.17.
\newline The entries marked with a dash are not applicable for $M_2=3$.} 
 \begin{tabular}{c|c|ccc}
$am_{\alpha}$&Case Q&Case A&Case B&Case C\\
$ v_i^{\alpha}$&$3\times 3$ $T_{{\rm min}}=4$&$3\times 3$ $T_{{\rm min}}=4$&$4\times 4$ $T_{{\rm min}}=3$&$4\times 4$ $T_{{\rm min}}=4$\\ \hline
$am_{1}$& 0.8721(19)&0.8580(11) &0.8340(40)& 0.8280(88)\\
$am_{2}$&1.263(13)&1.2267(51)&1.166(11)&1.138(35) \\
$am_{3}$&1.94(30)&1.93(13)&1.632(42)&1.52(12) \\
$am_{4}$&--&--&1.889(58)&1.85(18)\\
$v^1_{{\rm L}}$&0.4847(56)&0.4344(31)&0.3757(99)&0.359(24)\\
$v^1_{{\rm F}_1}$&1.519(10)&1.3779(70)&1.227(26)&1.181(65)\\
$v^1_{{\rm F}_2}$&0.8402(38)&0.8008(21)&0.731(12)&0.711(31)\\
$v^2_{{\rm L}}$&0.816(16)&0.8405(65)&0.801(12)&0.757(55)\\
$v^2_{{\rm F}_1}$&0.644(49)&0.874(22)&1.185(44)&1.260(83)\\
$v^2_{{\rm F}_2}$& --0.251(33)&--0.115(12)&0.169(47)&0.26(12)\\
$v^3_{{\rm L}}$&--0.28(22)&--0.348(97)&--0.459(29)&--0.534(85)\\
$v^3_{{\rm F}_1}$& 2.2(1.4)&2.34(63)&1.36(16)&0.86(53)\\
$v^3_{{\rm F}_2}$&--1.13(81)&--0.84(39)&0.56(19)&0.61(25) \\
$v^4_{{\rm L}}$& -- &--&0.000(69)&0.05(19)\\
$v^4_{{\rm F}_1}$&--&--&--0.45(33)&--0.73(70)\\ 
$v^4_{{\rm F}_2}$&--&--&1.89(22)&1.64(78)\\ \hline
$n_{2,{\rm data}}$&54&48&54&48\\
$n_{2,{\rm param}}$&12&12&16\{7\}&16\\
$n_{2,{\rm dof}}$&42&36&38\{45\}&32 \\
$\chi^2/n_{2,{\rm dof}}$&0.65&4.27&0.16\{0.13\}&0.14\\ \hline
$\Delta m_{21}$&0.391(13)&0.369(5)&0.332(12)&0.310(36)\\
&[0.322(11)]&&&\\
$\Delta m_{31}$&1.07(30) &1.07(13)&0.798(42)&0.69(12) \\
&[0.88(24)]&&&\\
$\Delta m_{41}$&--       &--      &1.05(6)  &1.02(18) \\
\end{tabular}
\label{Tablec2fit}
\end{table}

\begin{table}
\caption{The ratio $R(T)$ defined by Eq.~\protect\ref{ratio} and 
having the symmetry $R(T)=R(T+1)$ in Eq.~\protect\ref{sym},
when $T$ is odd. Here $R(T)$ is calculated for
$C_{3,{\rm F}_1{\rm F}_1}$ --- the dominant
correlation.}
 
\begin{tabular}{ccccc}
T&$R(T,r=1)$&$R(T,x=1,y=1)$&$R(T,x=1,y=1,z=1)$&$R(T,x=3,y=4)$\\ \hline
5&1.0222&0.958&0.890&  0.611 \\
6&1.0223&0.955&0.882& 0.585 \\
7&0.9536&0.919&0.876&0.598\\
8&0.9507&0.913&0.874&0.544\\
9&0.9166&0.879&0.844& 0.590   \\
10&0.9133&0.857&0.839& 0.524\\
\end{tabular}
\label{symt}
\end{table}

\begin{table}

\caption{Estimates of the charge radial distributions [$x^{11}(r)$] for
the ground state.
\newline Separable case (S): $M_2=4$ in $C_2$, $M_3=3$ in C(3).
$T_{2,{\rm min}}=3$ and $T_{3,{\rm min}}=6,8$.
\newline Separable case (S): $M_2=4$ in $C_2$, $M_3=3$ in C(3).
$T_{2,{\rm min}}=4$ and $T_{3,{\rm min}}=8$.
\newline Non-separable case (NS): $M_2=M_3=4$ in $C_2$ and $C_3$ but
$x^{ij}=0$, when neither $i$ nor $j$ is 1. $T_{2,{\rm min}}=3$ and $T_{3,{\rm min}}=6,8$.
\newline The column labelled Best Estimate is a summary of the previous
5 columns.
\newline The column labelled Approx. Estimate is from
Eq.~\protect\ref{xij} for $M_2=3$ ---
Case A in Table \protect\ref{Tablec2fit}.
\newline The line marked as 'ave' is the weighted mean
$[6x^{11}(3,0,0)+24x^{11}(2,2,1)]/30$  }
 \begin{tabular}{c|lll|lll|ll|ll}
\multicolumn{4}{c|}{}&S&S&S&NS&NS&Best&Approx.\\
\multicolumn{4}{c|}{$T_{2,{\rm min}}$}&3&3&4&3&3&Estimate&Estimate\\
\multicolumn{4}{c|}{$T_{3,{\rm min}}$}&6&8&8&6&8&&\\ \hline
r&x   &  y &z&   &&&&& \\ \hline
0.00&0&0&0 &0.0378(2)  &0.0377(4)  &0.0349(5)  &0.0428(3)  &0.0401(8)  &0.038(4)   &0.047(2)  \\
1.00&1&0&0 &0.01214(10)&0.0127(2)  &0.0112(3)  &0.01343(15)&0.0135(4)  &0.0125(15) &0.015(1)  \\
1.41&1&1&0 &0.00704(8) &0.0075(2)  &0.0068(2)  &0.00831(10)&0.0091(3)  &0.0080(15) &0.009(2)  \\
1.73&1&1&1 &0.00501(8) &0.0052(2)  &0.0048(2)  &0.00594(11)&0.0063(3)  &0.0055(7)  &0.0060(3) \\
2.00&2&0&0 &0.00749(9) &0.0072(2)  &0.0070(3)  &0.00627(11)&0.0068(3)  &0.0070(5)  &0.0071(5) \\
2.24&2&1&0 &0.00476(6) &0.00470(13)&0.0045(2)  &0.00464(6) &0.0047(2)  &0.0047(2)  &0.0046(3) \\
2.83&2&2&0 &0.00307(7) &0.0031(2)  &0.0031(2)  &0.00299(7) &0.0031(2)  &0.0032(2)  &0.0028(2) \\
3.00&3&0&0 &0.00329(10)&0.0034(3)  &0.0035(3)  &0.00295(11)&0.0035(3)  &0.0034(4)  &0.0032(5) \\
3.00&2&2&1 &0.00221(6) &0.00216(15)&0.0021(2)  &0.00215(6) &0.0019(2)  &0.00215(15)&0.0021(2) \\
3.00&\multicolumn{3}{c|}{ave}
           &0.00243(6) &0.00241(13)&0.0023(2)  &0.00231(5) &0.0022(2)  &0.00240(14)&0.0023(2) \\
3.16&3&1&0 &0.00234(6) &0.00244(14)&0.0025(2)  &0.00226(6) &0.0026(2)  &0.0025(3)  &0.0024(2) \\
3.61&3&2&0 &0.00164(6) &0.00178(13)&0.0019(2)  &0.00166(5) &0.00188(14)&0.0018(2)  &0.0017(2) \\
4.00&4&0&0 &0.00141(10)&0.0015(3)  &0.0016(4)  &0.00126(10)&0.0014(3)  &0.0015(3)  &0.0012(2) \\
4.12&4&1&0 &0.00095(5) &0.00105(13)&0.0011(2)  &0.00092(5) &0.00115(15)&0.0011(2)  &0.0009(1) \\
4.24&3&3&0 &0.00082(8) &0.0010(2)  &0.0010(3)  &0.00085(7) &0.0011(2)  &0.0010(3)  &0.0008(2) \\
4.47&4&2&0 &0.00074(5) &0.00075(12)&0.0007(2)  &0.00076(5) &0.00068(15)&0.00070(15)&0.0005(2) \\
5.00&4&3&0 &0.00067(6) &0.00080(10)&0.0009(2)  &0.00078(5) &0.00115(14)&0.0009(3)  &0.0008(2) \\
5.10&5&1&0 &0.00046(6) &0.00053(14)&0.0006(2)  &0.00050(5) &0.0007(2)  &0.0006(2)  &0.0005(2) \\
5.39&5&2&0 &0.00033(8) &0.00044(14)&0.0005(2)  &0.00036(5) &0.0006(2)  &0.0005(2)  &0.0003(1) \\
5.83&5&3&0 &0.00020(5) &0.00026(12)&0.0003(2)  &0.00027(5) &0.0005(2)  &0.00035(15)&0.0004(2) \\
\hline
\end{tabular}
\label{Tablechden}
\end{table}

\begin{table}
\caption{Best estimates of radial distributions of charge and matter
distributions involving also excited states. The column labelled
$x^{11}$~(AE) is an approximate estimate from Eq.~\protect\ref{xij}
for $M_2=3$ --- Case A in Table \protect\ref{Tablec2fit}. The entry
'--' implies that a reasonable signal could not be obtained and the
entry 'ave' has the same definition as in
Table~\protect\ref{Tablechden}.
}
 \begin{tabular}{c|lll|lll}
\multicolumn{1}{c}{}&\multicolumn{3}{c|}{Charge}&
\multicolumn{3}{c}{Matter}\\ \hline

r&$x^{12}$ &$x^{13}$& $x^{22}$&$x^{11}$&$x^{11}$~(AE)&$x^{12}$\\ \hline
$0.00$ &$0.079(4)$    &$-0.050(2)$  &$0.18(2)$    &$0.045(3)$ &$0.058(2)$ &$0.090(5)$  \\
$1.00$ &$0.0160(5)$   &$0.015(2)$   &$0.021(3)$   &$0.0133(5)$&$0.016(1)$ &$0.009(1)$  \\
$1.41$ &$0.0045(15)$  &$0.010(2)$   &$0.003(1)$   &$0.0073(7)$&$0.0076(6)$&$-0.001(1)$ \\
$1.73$ &$0.0010(10)$  &$0.007(2)$   &$0.005(4)$   &$0.0049(5)$&$0.0045(6)$&$-0.0030(7)$\\
$2.00$ &$0.0020(5)$   &$0.0005(15)$ &$0.0005(4)$  &$0.006(1)$ &$0.0054(5)$&$-0.003(1)$ \\
$2.24$ &$0.0000(3)$   &$0.0022(5)$  &$0.0000(1)$  &$0.0038(4)$&$0.0032(3)$&$-0.0030(7)$\\
$2.83$ &$-0.0010(3)$  &$-0.0005(10)$&$0.0003(2)$  &$0.0017(2)$&$0.0013(3)$&$-0.0015(3)$\\
$3.00$ &$-0.0009(4)$  &$-0.0018(15)$&$0.0003(2)$  &$0.0014(3)$&$0.0013(3)$&$-0.0010(3)$\\
$3.00$ &$-0.0007(2)$  &$0.0010(10)$ &$0.0003(2)$  &$0.0016(2)$&$0.0009(2)$&$-0.0018(5)$\\
ave    &$-0.0007(2)$  &$0.0004(9)$  &$0.0003(2)$  &$0.0016(2)$&$0.0010(2)$&$-0.0016(4)$\\
$3.16$ &$-0.0009(2)$  &$-0.0010(10)$&$0.0003(2)$  &$0.0009(2)$&$0.0009(3)$&$-0.0005(5)$\\
$3.61$ &$-0.0010(2)$  &$-0.0010(5)$ &$0.0006(2)$  &$0.0007(2)$&$0.0005(3)$&$-0.0008(4)$\\
$4.00$ &$-0.0007(2)$  &$-0.0018(12)$&$0.0005(3)$  &--&--&--\\
$4.12$ &$-0.00045(15)$&$-0.0010(8)$ &$0.0003(2)$  &--&--&--\\
$4.24$ &$-0.00050(15)$&$-0.0011(8)$ &$0.0005(3)$  &--&--&--\\
$4.47$ &$-0.00040(15)$&$-0.0000(5)$ &$0.00025(15)$&--&--&--\\
$5.00$ &$-0.0007(2)$  &$-0.0010(5)$ &$0.00065(15)$&--&--&--\\
$5.10$ &$-0.00040(15)$&$-0.0008(4)$ &$0.0004(2)$  &--&--&--\\
$5.39$ &$-0.0003(2)$  &$-0.0005(3)$ &$0.0003(2)$  &--&--&--\\
$5.83$ &$-0.0003(2)$  &$0.0001(3)$  &$0.0001(2)$  &--&--&--\\ \hline
\end{tabular}
\label{Tablechdenex}
\end{table}
\begin{table}

\caption{The charge and matter sum rules in the ground state $x^{11}$ and
between  excited states.
\newline The symbols N and NS are as in Table~\protect\ref{Tablechden}.
\newline Q refers to the quenched results (case 3) in
Ref.~\protect\cite{G+K+P+M}.
\newline The entries marked with a dash are not applicable for $M_3=3$. }
 \begin{tabular}{c|lll|llll}
&S&S&S&NS&NS&NS&Q-NS\\
$T_{2,{\rm min}}/T_{3,{\rm min}}$&(3/6)&(3/8)&(4/8)&(3/6)&(3/7)&(3/8)&(3/8)\\
$x^{\alpha \beta}$& &   &&&&& \\ \hline
Charge&&&&&&&\\ \hline
$x^{11}$ &$1.12(3)$ &$1.26(6)$  &$1.26(8)$  &$1.29(4)$  &$1.36(6)$  &$1.42(11)$ &$1.41(5)$ \\
$x^{12}$ &$0.09(3)$ &$-0.10(9)$ &$-0.05(12)$&$-0.28(7)$ &$-0.40(12)$&$-0.5(2)$  &$-0.5(2)$ \\
$x^{13}$ &$0.02(9)$ &$-0.3(4)$  &$-0.1(4)$  &$0.09(11)$ &$-0.2(2)$  &$-0.3(6)$  &$0(1)$    \\
$x^{14}$ &--        &--         &--         &$-0.1(2)$  &$-0.4(4)$  &$-0.5(1.1)$&--        \\
$x^{22}$ &$0.008(5)$&$0.007(14)$&$0.002(9)$ &$0.8(2)$   &$0.9(3)$   &$1.1(6)$   &$0.9(9)$  \\
\hline
$\chi^2/n_{{\rm dof}}$&$2.52$&$0.65$&$0.63$&$0.31$&$0.13$&$0.09$&$0.26$\\
\hline
Matter&&&&&&&\\ \hline
$x^{11}$ &$0.66(5)$ &$0.85(14)$ &$1.0(2)$   &$0.73(8)$  &$0.83(11)$ &$1.1(2)$   &$0.38(15)$\\
$x^{12}$ &$-0.15(6)$&$-0.5(2)$  &$-0.6(3)$  &$-0.27(13)$&$-0.4(2)$  &$-0.8(4)$  &$-0.1(5)$ \\
$x^{13}$ &$-0.4(2)$ &$-1.7(1.0)$&$-1.6(9)$  &$-0.3(2)$  &$-0.9(4)$  &$-2.7(1.2)$&$-1(5)$   \\
$x^{14}$ &--        &--         &--         &$-0.2(3)$  &$-0.8(7)$  &$-3.2(2.2)$&$-$       \\
$x^{22}$ &$0.04(3)$ &$0.3(2)$   &$0.4(4)$   &$0.2(3)$   &$0.2(6)$   &$0.2(1.2)$ &$1(2)$    \\
\hline
$\chi^2/n_{{\rm dof}}$&$0.46$&$0.29$&$0.25$&$0.31$&$0.25$&$0.12$&$0.35$\\
\end{tabular}
\label{Chsumrule}
\end{table}

\begin{table}
\caption{Fits to the charge lattice data with  exponential, Yukawa
 and Gaussian forms.
\newline Column 2E refer to the  exponential form 
$\sum_{0,1} a_i^{{\rm E}}\exp(-r/r_i^{{\rm E}})$.
\newline Column 2Y refer to the  Yukawa form
$\sum_{0,1} a_i^{{\rm Y}}\exp(-r/r_i^{{\rm Y}})/r$.
\newline Column 2G  refer to the  gaussian form
$\sum_{0,1} a_i^{{\rm G}}\exp(-(r/r_0^{{\rm G}})^2)$.
\newline Also fits to the lattice data with  the lattice Yukawa (2LY),
exponential (2LE) Gaussian (2LG)  forms from Eqs.~\protect\ref{LY},
 \protect\ref{LE} and  \protect\ref{LG}.
\newline $I_{{\rm E}}=8\pi\sum_{0,1} a_i^{{\rm E}}(r_i^{{\rm E}})^3$,
$I_{{\rm Y}}=4\pi \sum_{0,1}a_i^{{\rm Y}}(r_i^{{\rm Y}})^2$, 
 $I_{{\rm G}}=\pi^{3/2}\sum_{0,1} a_i^{{\rm G}}(r_i^{{\rm G}})^3$
\newline are the spacial integrals of these functions.
\newline The entries marked as $^*$ are fixed in the minimization.}
\begin{tabular}{c|cc|cc|cc}
Form($F$)&2E&2LE&2Y&2LY&2G&2LG\\ \hline
$a_0^{{\rm F}}$&0.0250(22)&0.0245(21)&0.069(10)&0.066(3)&0.0086(5)&0.0075(7)\\
$r_0^{{\rm F}}$&1.36(6)&1.37(6)  &1.70(6)&1.78(7)&2.92(9)&3.07(12)\\
$a_1^{{\rm F}}$&0.013(5)&0.373$^*$&--0.073$^*$&--0.060$^*$&0.0293(41)&0.019(4)\\
$r_1^{{\rm F}}$&0.2(4)&0.116(15)&1.00(17)&1.12(11)&0.75(7)&0.99(15)\\ 
$I^{{\rm F}}$&1.6(3)&1.6(3)&1.6(5)&1.7(3)&1.3(1)&1.3(2)\\ \hline
$\chi^2/n_{{\rm dof}}$&1.32&0.81&1.33&0.94&1.47&0.93\\
\end{tabular}
\label{LFits}
\end{table}

\newpage
\begin{table}
\caption{Fits to the matter lattice data with  exponential, Yukawa
 and Gaussian forms. In 2LY$^{\dagger}$ the $r=0$ data point is not fitted. 
Other notation as in Table~\protect\ref{LFits}.}
\begin{tabular}{c|cc|cc|cc}
Form (F)&2E&2LE&2Y&2LY$^{\dagger}$&2G&2LG\\ \hline
$a_0^{{\rm F}}$&0.036(20)&0.0345(26)&0.178(12)&0.186(23)&0.0105(12)&0.0101(16)\\
$r_0^{{\rm F}}$&0.91(13)&0.938(39)  &1.04(12)&0.92(6)&2.11(9)&2.14(11)\\
$a_1^{{\rm F}}$&0.009(20)&0.561$^*$&--0.168$^*$&--0.207$^*$&0.0346(32)&0.034(10)\\
$r_1^{{\rm F}}$&0.45(55)&0.099(11)&0.90(18)&0.71(12)&0.71(4)&0.70(12)\\ 
$I^{{\rm F}}$&0.7(5)&0.73(10)&0.7(9)&0.7(6)&0.62(10)&0.62(13)\\ \hline
$\chi^2/n_{{\rm dof}}$&1.80&1.16&1.74&1.11&1.11&1.00\\
\end{tabular}
\label{LFitsm}
\end{table}

\begin{table}
\caption{Fits to the separable function $y_1(r)$ defined as 
$y_1(r)=(x^{11})^{1/2}$, where the $x^{11}$ are the charge density values
labelled as Best Estimate in \protect\ref{Tablechden}.
Notation as in Table~\protect\ref{LFits}.}
\begin{tabular}{c|cc|cc|cc}
Form (LFS)           &2ES     &2LES&2YS      &2LYS&2GS&2LGS      \\ \hline
$a_0^{{\rm LFS}}$     &0.159(7)&0.156(8)&0.47(3)&2.4310(60)&0.092(3)&0.083(4)\\
$r_0^{{\rm LFS}}$     &2.73(12)&2.59(9)&4.40(21)&2.86(11)&4.19(12)&4.44(17) \\
$a_1^{{\rm LFS}}$     &0.036(13)&0.0422$^*$&--0.49$^*$&--2.4196$^*$&0.102(11)
&0.065(6)\\
$r_1^{{\rm LFS}}$     &0.2(5)&0.41(17)&1.66(22)&2.43(10)&0.85(9)&1.24(15)   \\  \hline
$\chi^2/n_{{\rm dof}}$ &1.32&1.11&1.33    &1.11&1.75&1.07     \\
\end{tabular}
\label{LFitswave}
\end{table}

\begin{table}
\caption{Fits to the separable function $y_1(r)$ defined as 
$y_1(r)=(x^{11})^{1/2}$, where the $x^{11}$ are the matter density values
 in \protect\ref{Tablechdenex}. In 2LY$^{\dagger}$ the $r=0$ data point
is not fitted.   
Other notation as in Table~\protect\ref{LFits}.}
\begin{tabular}{c|cc}
Form (LFS)     &2YS      &2LYS$^{\dagger}$      \\ \hline
$a_0^{{\rm LFS}}$     &0.75(4)&0.74(4)\\
$r_0^{{\rm LFS}}$     &2.39(27)&2.25(18) \\
$a_1^{{\rm LFS}}$     &--0.728$^*$    &--0.774$^*$\\
$r_1^{{\rm LFS}}$     &1.52(28)&1.36(20)   \\  \hline
$\chi^2/n_{{\rm dof}}$ &1.84    &1.25   \\
\end{tabular}
\label{LFitswavem}
\end{table}

\newpage

\begin{table}
\caption{The value of $r_0$ from fits to the Fourier transform of the 
charge density  $x^{11}(\bf r)$ as defined in Eq.~\protect\ref{FT}.
\newline  Row A: $r_0$ directly from Tables~\protect\ref{LFits}
and \protect\ref{LFitswave}.
\newline  Row B: $r_0$ extracted from the Fourier Transform of
$x^{11}(\bf r)$ when these are expressed in terms of the
analytic expressions in Eqs.~\protect\ref{LY}, \protect\ref{LE} and
\protect\ref{LG} with the parameters in Tables~\protect\ref{LFits} and
\protect\ref{LFitswave}. Option (i) is used for defining the $\chi^2$.
\newline  Row C: Same as Row A but using Option (ii).   
\newline Row D: The $x^{11}(\bf r)$ are the same as in row B and C, except
that the data from Table~\protect\ref{Tablechden} is used wherever
possible --- Option (i) used.
\newline Row E: Same as Row D but using Option (ii).
\newline The entry '--' means a sensible solution could not be found} 
\begin{tabular}{c|cc|c}
\multicolumn{1}{c}{Extract with}&\multicolumn{2}{c|}{2LY}&
\multicolumn{1}{c}{2LE}\\ \hline 
\multicolumn{1}{c|}{A}
&\multicolumn{2}{c|}{1.78(7)}
&\multicolumn{1}{c}{1.37(6)}\\ \hline
Form&2LY&2LE&2LE \\ \hline
B&1.78(24)&1.75(24)&1.38(22) \\
C&1.78(9)&2.5(6)&1.38(15)\\
D&1.78(20)&1.75(22)&1.39(23)\\
E&3.3(1.0)&--&1.41(16)\\
\end{tabular}
\label{polefit}
\end{table}

\begin{table} 
\caption{Estimates of the charge sum rule from the  ${\bf k=0}$
component of Eq.~\protect\ref{FT}. The ${\bf r}$ summation is truncated
in two ways --- cubic(C) cut-off, where $|x|, \ |y|$ and $|z|$ are all
$\le r_c$ and the spherical(S) cut-off, where only values of $x, \ y$ and
$z$ with $x^2+y^2+z^2\le r_c^2$ are kept in the summation. The columns
labelled $I^{2{\rm E}}$ and $I^{2{\rm Y}}$ are the integrals from 0 to
$r_c$ of the functions 2E and 2Y in Table~\protect\ref{LFits}.
The row labelled
None means the ${\bf r}$ summation is over the whole $L^3$ lattice.  } 
 \begin{tabular}{c|ccc|ccc}
\multicolumn{1}{c|}{}
&\multicolumn{2}{c}{2LE+data}&\multicolumn{1}{c|}{2E}
&\multicolumn{2}{c}{2LY+data}&\multicolumn{1}{c}{2Y}\\
$r_c$&C&S&$I^{2{\rm E}}$&C&S&$I^{2{\rm Y}}$\\ \hline
3&0.95&0.62&0.63&0.95&0.62&0.58\\
4&1.20&0.87&0.90&1.20&0.87&0.85\\
5&1.37&1.11&1.10&1.38&1.11&1.06\\
6&1.49&1.30&1.23&1.51&1.30&1.21\\
7&1.58&1.42&1.31&1.61&1.42&1.37\\
None&1.62&1.62&1.6(3)&1.65&1.65&1.6(5)\\
\end{tabular}
\label{cutoff}
\end{table}

\newpage

\begin{table}
\caption{This is a continuation of Table~\protect\ref{cutoff}
for the spherical cutoff case (S) containing the available data.
Here single
exponential (E), Yukawa (Y) and gaussians (G) tails are included for
$r>r_c$. The parameters of these tails are obtained from data with 
$r\ge 3$. The combined functions (2LE+E, 2LY+Y,
2LE+G and 2LY+G) are integrated using the cubic cutoff with $r=7$. }
\begin{tabular}{c|cccc}
$r_c$&2LE+E&2LY+Y&2LE+G&2LY+G\\ \hline
3&1.42&1.46&1.30&1.30\\
4&1.40&1.45&1.28&1.28\\
5&1.41&1.46&1.29&1.29\\ 
\end{tabular} 
\label{Tails}
\end{table}  

\begin{table}
\caption{Check on the rotation invariance of 
$R(r)=\langle C_{3,{\rm F}_1{\rm F}_1}(T, \ r)\rangle /\langle
C_{2,{\rm F}_1{\rm F}_1}(T)\rangle $
for $r=3$ versus $x=2, \ y=2, \ z=1$ and $r=5$ versus $x=3, \ y=4$.
All entries  should be multiplied  by $10^{-4}$ }
\begin{tabular}{c|cc|cc}
T&$R(r=3)$&$R(x=2,y=2,z=1)$&
$R(r=5)$&$ R(x=3,y=4)$\\ \hline
4&23.0(4)&11.6(2)&1.2(4)&1.2(2)\\
5&24.4(4)&13.4(3)&1.7(4)&1.6(2)\\
6&25.4(6)&15.3(4)&2.0(5)&2.2(3)\\
7&25.7(7)&16.2(4)&1.1(6)&2.8(3)\\
8&26.2(1.1)&17.7(0.7)&0.8(9)&3.9(5)\\
9&26.8(1.5)&17.4(0.9)&1.0(1.2)&5.0(7)\\
10&28.8(2.7)&17.4(1.5)&2.9(2.2)&7.0(1.3)\\
\end{tabular}
\label{rotv2}
\end{table}

\begin{table}
\caption{Lack of rotation invariance in the lattice Yukawa and
exponential forms in Eqs.~\protect\ref{LY} and ~\protect\ref{LE} for 
 $r=3$ versus $(x,y,z)=(2,2,1)$ and $r=5$ versus $(x,y)=(3,4)$. } 
\begin{tabular}{l|ccc}
Case&  $r=3$   &     $(2,2,1)$ &   Ratio \\ \hline
2LY(Table~\protect\ref{LFits})  & 0.002893   &    0.002566 &    0.89 \\
2LE(Table~\protect\ref{LFits})  & 0.002893   &    0.002573 &    0.89 \\ \hline
   &     $r=5$   &       $(3,4)$&      Ratio \\ \hline
2LY(Table~\protect\ref{LFits})  & 0.000709  &    0.000653  &  0.92 \\
2LE(Table~\protect\ref{LFits})  & 0.000706  &    0.000651  &  0.92 \\
\end{tabular} 
\label{rotvlye} 
\end{table} 
\newpage

\newpage 
  
\newpage

\newpage

\newpage

\newpage
  
\newpage

\end{document}